\documentclass{ws-book961x669}
\usepackage{ws-book-thm}      
\usepackage{ws-book-van}      

\begin{document}

%








\setcounter{page}{1}


\chapter{The odderon and BKP states in the Quantum Chromodynamics}
%
\begin{center}
M.A.Braun\\
Dep. of High Energy physics,
Saint-Petersburg State University,\\
198504 S.Petersburg, Russia\\ 
\vskip 0.5cm
G.P. Vacca\\
 INFN Sezione di Bologna, Via Irnerio 46, 40126, Bologna, Italy
 \end{center}
\newcommand\beq{\begin{equation}}
\newcommand\eeq{\end{equation}}
\newcommand\bea{\begin{eqnarray}}
\newcommand\eea{\end{eqnarray}}
\newcommand\lra{\leftrightarrow}
\def\oq{\omega(q)}
\def\oa{\omega(q_{1})}
\def\ob{\omega(q_{2})}
\def\eq{\eta (q)}
\def\ea{\eta (q_{1})}
\def\eb{\eta (q_{2})}
\def\ec{\eta (q'_{1})}
\def\ed{\eta (q'_{2})}
\def\fq{f (q)}
\def\fa{f (q_{1})}
\def\fb{f (q_{2})}
\def\fc{f (q'_{1})}
\def\fd{f (q'_{2})}
\def\ql{(q \cdot l)}
\def\qlp{(q \cdot l')}
\newcommand{\mbf}[1]{\mbox{\boldmath $#1$}}
\newcommand{\overright}{\mbf}
\newcommand{\bk}{\mbf{k}}
\newcommand{\bq}{\mbf{q}}
\newcommand{\bp}{\mbf{p}}
\newcommand{\bzero}{\mbf{0}}
\newcommand{\br}{\mbf{r}}
\newcommand{\bx}{\mbf{x}}
\newcommand{\by}{\mbf{y}}
\newcommand{\bz}{\mbf{z}}
\newcommand{\bR}{\mbf{R}}
\newcommand{\frho}{\mbox{\boldmath $\rho$}}
\newcommand{\psidag}{\psi^{\dag}}
\newcommand{\phidag}{\phi^{\dag}}
\newcommand{\deltadag}{\delta^{\dag}}
\newcommand{\chidag}{\chi^{\dag}}
\newcommand{\tildepsi}{\tilde{\psi}}
\newcommand{\tildepsidag}{\tilde{\psi}^{\dag}}



\include{graphics}

\centerline{\bf Abstract}

We discuss the odderon in the QCD and analyze the effects of running coupling introduced in a particular way
to preserve the  reggeization of the gluon within the leading logarithmic description. This idea is also applied to a family of 
BKP states with arbitrarily number of gluons in the planar limit. The numerical analysis shows that
contrary to the pomeron case where the leading states become discretized,
the odderon states still remain  a continuous family starting at intercept one. The following 
rapidity dependence of the amplitude is studied. For the pomeron-odderon system the relation between the descriptions in  the reggeized gluon (BFKL) framework and  the 
color dipole/CGC one is investigated.

\newpage
\section {Introduction}
The odderon has a long history in the High-Energy physics. It started in 1973 and its first 30 years of existence
are exposed in the comprehensive review ~\cite{ewerz}. So here we only mention some principal points and developments after 2003.
The odderon was born as a result of an abstract idea that "strong interactions are as strong as they can be" in the paper
by  L. Lukaszuk and B. Nicolescu ~\cite{lukanic} and was given a name "maximal odderon".

The famous Froissart theorem establishes a limitation on the behavior of the scattering amplitudes governed by
strong interactions as
\beq
|A(s,t)|<Cs\log^2s,\ \ C<\pi/m_{\pi}^2\simeq 62\ mbn.
\eeq
In fact the amplitude can be split into $C=+1$ and $C=-1$ parts (or signatures $\xi=\pm)$:
\beq
A^{(\pm)}(s)=\frac{1}{2}[ A(s)\pm A(-s) ]
\eeq
and the long-standing Pomeranchuk theorem asserted
\beq
\frac{A^-(s)}{A^+(s)}_{s\to \infty}\to 0\,,
\eeq
where the leading part of the $A^{(+)}$ and $A^{(-)}$ in the high energy limit is associated to the pomeron and to the odderon exchange, respectively.
The results following from the Regge theory and based on the leading  $\omega,\rho$-Regge trajectory in the $C=-1$ sector  fully
confirmed this rule. The   $\omega,\rho$ intercept is $\alpha_{\rho,\omega}(0)\simeq 1/2$ so that the corresponding reggeon
interchange  leads to the amplitude $A^{(-)}\sim\sqrt{s}$, which is fully confirmed by the experiment.

The idea of the maximal odderon proclaimed that both amplitudes $A^{(\pm)}$ take its maximal value at $s\to\infty$, with the only
difference that
$A^{(+)}(s)$ is positive  imaginary and $A^{(-)}(s)$ real:
\beq
A^{(+)}(s)\sim is(\log^2s-i\pi\log s),\ \ A^{(-)}\sim s(\ln^2s-i\pi\log
s),
 \ \ s>>m^2.
 \label{eq0}
\eeq
Let us briefly recall  some simple and evident properties of the maximal
odderon amplitude (\ref{eq0}) (at $t=0$). As mentioned it is mostly real. Its
imaginary part is
proportional to $s\log s$, which corresponds to the difference between
the particle-particle and particle-antiparticle cross-sections growing
like $\log s$:
\beq
\sigma_{a\bar{b}}(s)-\sigma_{ab}\sim \log s, \ \ s>>m^2.
\eeq
Since the total cross-section for each of the two reactions grows like
$\log^2s$, the Pomeranchuk theorem is fulfilled in the sense
\beq
\frac{\sigma_{a\bar{b}}(s)}{\sigma_{ab}(s)}\rightarrow 1,\ \
s\rightarrow\infty.
\eeq
The most striking feature of the asymptotic odderon is a nonvanishing
ratio  of the real and imaginary parts of the amplitudes:
\beq
\frac{{\rm Re} A(s)}{{\rm Im}A(s)}\sim const, \ \ s>>m^2.
\eeq
In the course of time and depending on the experimental situation
these properties served as a dominant motive for and against the
introduction of
the odderon. We shall turn to the experimental evidence for the
odderon in the  end of this section.

In an attempt to realize this picture in the complex angular $j$-plane the authors
were obliged to introduce rather exotic singularities. Addressing the interested author to the review \cite{ewerz} to
see the details, we only mention that typically higher order poles at $j=1$
were introduced to generate the desired $\log(s)$ dependence. The authors had also to struggle against the
singularities which might appear in the physical amplitude, since for $A^{(-)}$ the point $j=1$ is physical.

With the advent of the Quantum Chromodynamics the odderon acquired a new life. In the lowest order there one finds both
$C=\pm 1$ amplitudes on equal footing, the $C=+1$ one coming from the exchange of two gluons and the $C=-1$
from the exchange of three gluons. Both amplitudes are linear in $s$, so that their behavior at $s\to\infty$ is
similar and corresponds to a pole in the $j$ plane at $j=1$. Ever since the theoretical efforts were directed to
find corrections to this simple results in higher orders.  Remarkable success has been achieved in the so-called
leading-log approximation, in which terms of the order$(\alpha_s\ln(s))^n$ are summed at high $s$ and small $\alpha_s$.
A well-known BFKL equation was set up  and solved to describe the evolution of the pomeron as early as 1975
~\cite{bfkl1,bfkl2} and its generalization to the next-to leading order in was derived in 1995 ~\cite{lipfadin, ciafa}.
The vertex for splitting of
the pomeron in two was constructed in 1995 in two approaches, the dispersion one, following the derivation of the BFKL equation
in ~\cite{barwue} and in the approach based on the color dipole picture in ~\cite{mueller}. On the basis of this vertex
equations for the scattering of a small object on the nucleus were proposed by I.I.~Balitsky and Yu.~Kovchegov (the BK equation)
~\cite{bal, kov}, later generalized to CGC-JINWLK equation (see e.g. ~\cite{iancu}). Analogous developments were realized in the
BFKL-Bartels approach ~\cite{bra,blv1}.  Parallel to this the odderon was investigated both in the BFKL-Bartels and dipole
approaches. The equation for the odderon in the leading-log approximation was set up as one of the so-called BKP equations proposed
in ~\cite{bartels1,jaro,kwicz}. L.N.~Lipatov analyzed the conformal properties of the odderon  and related its equation to that
for the solvable chain of conformal spins ~\cite{lip}.  This allowed to R.A.~Janik and J.~Wosiek to obtain an explicit equation for the
odderon wave function (the JW odderon) and find its maximal intercept ~\cite{jw}
\beq
\alpha_O= 1-0.24717\frac{\alpha_sN_c}{\pi}\nonumber\,,
\eeq
where $\alpha_s=g_s^2/(4\pi)$.
Somewhat later a new solution  of the odderon equation was obtained by J.~Bartels, L.N.~Lipatov and G.P.~Vacca (the BLV odderon)
with $\alpha_O=1$ in which two of the three reggeons forming the odderon are fused into one~\cite{blv}. This simple odderon
was later also found in the dipole approach in ~\cite{kovch1}. On the basis of the latter the evolution equation was set up
for the combined pomeron-odderon system in ~\cite{hatta} and attempts at solving it were made in ~\cite{hatta, motyka},
although in a very simplified (one-dimensional) approach. As to the more complicated JW odderon,
the necessary pomeron-two odderon vertex
was found in ~\cite{barew}, although its application has not been attempted so far, given that these odderon states where shown to decouple from
some perturbative impact factors, while the BLV Odderon states are maximally coupled to them~\cite{bar2}.
Later, using the effective action for reggeized gluon constructed by L. N. Lipatov, the BKP kernel for three reggeized gluon systems in the NLL approximation was constructed~\cite{Bartels:2012sw}. This can be used to analyze the odderon both in QCD, but in the large $N_c$ limit, or in general supersymmetric extensions, in particular $N=4$ SYM for any $N_c$.
Then, starting from generalized bootstrap equations involving inelastic production amplitudes, which were conjectured~\cite{Bartels:2013yga} but have still to be proved,
a construction for a family of odderon solution at NLL was given. They are extensions of the BLV solution to the NLL approximation and have also intercept starting strictly at one~\cite{Bartels:2013yga}.

We stress that in all developments of the odderon theory  seminal contributions were made by L.N.~Lipatov, beginning from the
study of the BFKL pomeron and including the understanding of the conformal properties of the pomeron and odderon, the equivalence
of the odderon equation to the solvable chain of conformal spins and construction of the BLV odderon.

On the experimental side the history of the odderon is just as long but marked by much less clarity,  novelty and success.
Manifestations of the odderon can be present either in the processes realized by the purely  odderon exchange or in the
interference of the pomeron and odderon exchanges. The typical example of the first case is the production of a PS meson
(C=+1) in the interactions starting with the photon (C=-1) either real or virtual. Discovery of this process would be a
straightforward detection of the odderon. There were various theoretical estimates of the rate of this transition
e.g. for the process $\gamma \to\eta_c$ ~\cite{bar2,bar3} (see also the review ~\cite{ewerz} for earlier works).
Unfortunately they all gave  estimations for the cross-sections far  below the present experimental possibilities. More informative
have been attempts to see the interference of the pomeron and odderon exchanges primarily in the comparison of
$pp$ and $p\bar{p}$ elastic scattering, In fact this comparison was the basic motivation for the introduction of the
odderon in the first papers in which this notion was introduced. Already at that time (in the seventies) a significant  difference
between these two cross-sections was discovered.
 In particular the dip present in the $pp$ cross-section as a function of $t$ was not seen in the
$p\bar{p}$ cross-section. At comparatively low energies this difference could be satisfactorily explained by the exchange of the
well-known $C=-1$ reggeons $\omega-\rho$ with an intercept much below unity. However the difference did not disappear at higher
energies. The situation at present date can be seen from Fig. \ref{fig1} in which the latest data on the elastic $pp$ scattering data
from the TOTEM collaboration at 2.76 TeV ~\cite{totem} are compared to the $p\bar{p}$ cross-sections from D0 at 1.96 Tev
~\cite{d0}.
Remarkably the picture is nearly identical  to the similar one for the  experiments  at 53 GeV   presented in the review
~\cite{ewerz} in 2003.  So the difference between $pp$ and $p\bar{p}$, if indeed it exists, does not depend on energy once it is
high enough. This latest experiments renewed extensive discussion of the odderon contribution at high energies, with not, however,
conclusive arguments, especially in view of considerable experimental uncertainties ~\cite{late1,late2, late3}, not to consider still open theoretical issues.

In the present paper we try to cover some theoretical questions left untouched in the  previous studies, which may throw additional light on the odderon as the QCD object.
In the first place we consider interacting pomeron-odderon systems in QCD in the generalized leading logarithmic approximation
to discuss the link between the reggeized gluon BFKL approach and the findings in the dipole/CGC picture. In particular
we try to construct the equations for the combined evolution of the pomeron, and BLV odderon in the dense (nuclear) target in the BFKL-Bartels approach to compare with similar equations in the dipole/CGC picture~\cite{hatta}.
We shall be able to derive only the first of these equations: evolution of the pomeron in the presence
of the odderon field. The second equation for the odderon itself requires knowledge of the vertex for the transition of three
reggeons to five reggeons, the task not realized so far  and evidently beyond the scope of this study. We hope to return to this
problem in future.
We also discuss what one may think could be an effective useful description in the non perturbative regime and in the large transverse distance limit,
the so called Reggeon Field Theory developed by V.Gribov for which we review some details of the universal critical behavior.
Then we study the influence of the running coupling on the odderon and
higher BKP states, which was not given much attention in the past. We also present in this framework the numerical study for both the pomeron and the odderon spectrum.
\begin{figure}
\vspace*{30 pt}
\centerline{\includegraphics[width=0.6\textwidth, angle=270,bbllx=0pt,bblly=0pt,bburx=594pt,bbury=700pt]{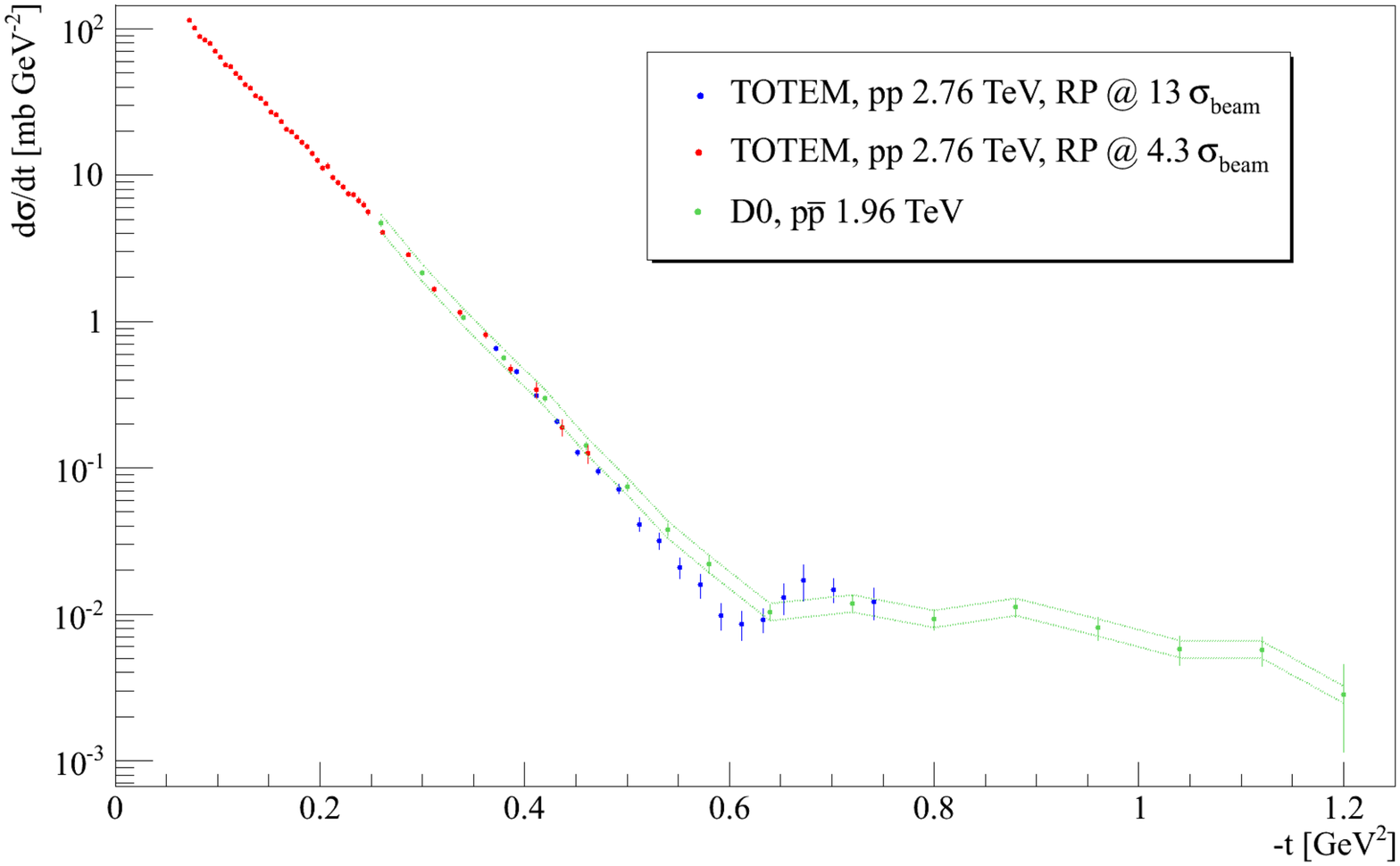}}
\caption{The differential cross sections ${\rm d}\sigma/{\rm d}t$ at $\sqrt{s}=2.76$~TeV measured by the
TOTEM experiment and the elastic $\rm p\bar{p}$ measurement of the D0 experiment at 1.96 TeV~\cite{d0}. The
		green dashed line indicates the normalization uncertainty of the D0 measurement.}
\label{fig1}
\end{figure}

\section{Interacting pomeron-odderon systems}
We start discussing the perturbative QCD behavior of pomeron-odderon system,
interactions which go in the direction of "unitarize" the theory in the sense of removing the violation of the Froissart bound.
Full evolution of effective interacting pomeron-odderon systems is too hard to be investigated in small x QCD, since it should involve also loops.
Nevertheless an effective tree level description has been used to study some features of the large rapidity evolution and propagation in a dense nuclear medium.
A simplified model, the so called Reggeon Field Theory is subsequently presented.
Computation with loops are at reach there and we give some recent perturbative
results for the critical properties.

\subsection{Fan diagrams for the pomeron-odderon system}
The distribution of gluons (and quarks) in hadronic scattering processes
depends on the rapidity under consideration and can be associated to phases characterized
by high densities of strongly interacting quanta.
Collisions with the  center of mass energy dominating over other momentum scales require
summation of contributions proportional to powers of $\log s$. This task was realized
by L.N.~Lipatov and collaborators in the so called BFKL approach.
The leading log (LL) and next-to-leading-log (NLL) approximations
resum $(\alpha_s \log s)^2$ and  $\alpha_s(\alpha_s \log s)^2$ contributions, respectively.
In the lowest order of perturbation theory it is sufficient to take into account only two reggeized gluons exchanged
in the colorless $t$ -channel. Then it was discovered that the cross sections grow as a power of energy in violation of the unitarity restrictions.
This  clearly showed the necessity to include more complicated structures into the $t$-channel,
larger numbers of reggeized gluons and transitions changing these numbers.
Of course  the final answer to the problem requires solving fully the Quantum Chromodynamics in the relevant kinematical conditions
(Regge kinematics), which does not seem realistic. So at present one has to rely on certain approximations
starting from the simplest BFKL approach. Their improvememnt can be realized by different simplified models based on
diagrams constructed from reggeized gluons in planar approximation or from pomeron diagrams  without loops.

Considerable progress has been achieved in the study  of the scattering of compact objects , quark-antiquark loops or "onia".
One starts from
the scattering amplitudes for the collision of two such objects. In the BFKL approach it can be factorized
 into impact factors (local in rapidity) and a BFKL pomeron Green function where the rapidity dependence is
encoded, both objects depending on the transverse momenta $\bq$ and $\bk$
\beq
 {\cal A}(Y,\bq) =i \int d^2\bk  \Phi_1(\bq,\bk) \, G_Y(\bq,\bk) \, \Phi_1(\bq,\bk) \,,
\eeq
The Green function $G_Y$ is associated to the evolution kernel later described in Section~\ref{sec:two_gluons}
restricting to a fixed coupling constant.
Including more $t$ channel gluons in the analysis and taking into account the change of their numbers
 has been a long lasting program started by J.~Bartels. Essentially the analysis was pushed up
to six $t$-channel reggeized gluons. It was based on the study of
multi-cut amplitudes $D_n$ for the transition of the quark-antiquark loop into $n$ reggeized gluons
(actually their multiple energy discontinuities). They can be related to the total cross sections  by
the AGK rules~\cite{Abramovsky:1973fm,Bartels:2005wa}.
A set of coupled equations has been constructed for $D_n$ with the following ingredients: impact factors $D_n^{0)}$
for $n$ reggeized gluons directly attached to the quark-antiquark loop,  the reggeized gluon trajectories, the $K_{n\to m}$
elementary transition vertices from $n$ to $m$ reggeized gluons and the so-called bootstrap relations.

As the outcome, the amplitude $D_n$ can be presented as a sum two contributions,  the so-called reggeized one $D_N^R$ for which
 evolution in rapidity is governed by a single pomeron Green function for two reggeized gluons $G_2$ (BFKL Green function),
and the other, irreducable one, $D_n^I$ which  evolves in rapidity via
Green functions $G_n$ for $n > 2$ reggeized gluons.
In particular for the $4$-gluon amplitude one finds
\beq
D_4=D_4^R+D_4^I \, \quad , D_4^I=\int G_4(Y-Y') V_{2\to4} D_2(Y') =G_4\otimes V_{2_\to 4} D_2 \,,
\label{4gluon}
\eeq
where $D_2(y')=G_2(y') D_2(0)$ and  the $V_{2\to4}$ vertex describes the  transition from $2$ to $4$ reggeized gluons
and $G_4$ is the Green function  for the general evolution of $4$ reggeized gluons.
Similarly for the $6$-gluon amplitude one finds ~\cite{barew}
\beq
D_6=D_6^R+D_6^I+D_6^E
\label{6gluon}
\eeq
where
\beq
 D_6^I=\int G_6(Y-Y') V_{2\to6} D_2(Y') =G_6\otimes V_{2_\to 6} D_2 \,.
 \label{D6I}
\eeq
It is remarkable that in the vertex  $V_{2\to 6}$
a term appears with a colour structure containing a product of two coefficients $d_{abc}$.
It  describes transition of a pomeron into two odderons.
The last term
\beq
D_6^E=\sum G_4 \otimes V_{2\to4} G_4 \otimes V_{2\to4} D_2+D^{extra}
\label{6gluonextra}
\eeq
contains contributions related to a double $2\to4$ splitting including the BKP evolution (fan structure) and finally another term with a peculiar tensor color structure which was not completely analyzed~\cite{barew}.

In an alternative, dipole or CGC approach, started by A.H.~Mueller with collaborators, the scattering of an "onium"
on a dense nuclear target was described (in the large $N_C$ limit) by the Balitski-Kovchegov evolution equation ~\cite{bal,kov}.
From the viewpoint of the BFKL-Bartels
approach this equation is equivalent to summation of fan diagrams made of BFKL green functions and triple pomeron vertices
$V_{2\to 4}$~\cite{Bartels:2004ef}.

From the start in the BFKL theory apart from the pomeron also the odderon was introduced as a compound state of three reggeized gluons with a color
factor $d_{abc}$. In its general form its wave function depends on the three gluon momenta (or coordinates). The equation for this wave function was
studied by L.N.~Lipatov ~\cite{lip} and, as mentioned in the Introduction, later solved by Janik and Woiscek (JW-odderon) who showed that
its ground state has an intercept smaller than unity. So the JW-odderon generates
cross-sections vanishing at high energies.

However later a different odderon state was found, which essentially depends on only two coordinates with a
fusing pair of the three gluons, the BLV-odderon ~\cite{blv}. Its wave function is essentially the antisymmetric pomeron wave function
whose ground state has the intercept exactly equal to one. Contribution from the BLV-odderon remains constant at high energies and
so dominate over the JW-odderon. Moreover some leading order impact factors are known to decouple from JW-odderon while couple maximally with the BLV-odderon~\cite{bar2}.
Explicitly the BLV-odderon eigenstate $\Psi$ is written in terms of the odd pomeron eigenstate $\psi$ as
\beq
\Psi= {\cal N} \hat{S} \psi
\label{blvodd}
\eeq
where $\hat{S}$ is an operator acting on 2-gluon
states and with values on the 3-gluon states, which performs an
antisymmetrization in the 2 incoming gluons, splits the first of them
in two and sums over the cyclic permutations of the outgoing gluons:
\beq
\hat{S}(1,2,3|1',2')\phi(1',2')=\frac{1}{2}\sum_{(123)}
[\phi(12,3)-\phi(3,12)] \, .
\label{hats}
\eeq
and ${\cal N}$ is a normalization factor.
One finds
\beq
\langle \Psi | \Psi \rangle = c_o  \langle \psi | {\cal N}^2 H_{12} | \psi \rangle\,, \quad
\eeq
where $c_o=3g_s^2 (N_c^2-4)/N_c^2$.
If one wants to have
the same normalized scalar product for $\Psi$ as for $\psi$ then
one has to choose ${\cal N}= ( c_o  H_{12})^{-1/2}$.

Somewhat later this odderon was also found in the dipole approach ~\cite{kovch1} also associated
with the antisymmetric
pomeron state. Subsequently this dipole odderon was included into the fan diagrams
giving rise to the evolution equations for the combined pomeron plus odderon evolution.
Denoting the pomeron and odderon density fields as  $N$ and $O$ respectively,
the equations are ~\cite{hatta}

\bea
\frac{d}{dY}N_{\overright{x},\overright{y}}=&{}&\bar{\alpha}_s
\int \frac{d^{2}z}{2\pi }\,\frac{\left| x-y\right| ^{2}}{\left|
x-z\right| ^{2}\left| y-z\right| ^{2}} \times\nonumber\\
&{}& \left( N_{\overright{x},%
\overright{z}}+N_{\overright{y},\overright{z}}-N_{%
\overright{x},\overright{y}}-N_{\overright{x},\overright{%
z}}N_{\overright{y},\overright{z}}+O_{\overright{x},\overright{%
z}}O_{\overright{y},\overright{z}}\right) \,.
\label{BKpom}
\eea

\bea
\frac{d}{dY}O_{\overright{x},\overright{y}}=&{}&\bar{\alpha}_s
\int \frac{d^{2}z}{2\pi }\,\frac{\left| x-y\right| ^{2}}{\left|
x-z\right| ^{2}\left| y-z\right| ^{2}} \times\nonumber\\
&{}& \left( O_{\overright{x},%
\overright{z}}+O_{\overright{y},\overright{z}}-O_{%
\overright{x},\overright{y}}-O_{\overright{x},\overright{%
z}}N_{\overright{y},\overright{z}}-N_{\overright{x},\overright{%
z}}O_{\overright{y},\overright{z}}\right) \,.
\label{BKodd}
\eea

Here we show that at least the first of these equations can be derived in the BFKL-Bartels framework
on the basis of the known transitional vertex $V_{2\to 6}$ constructed in ~\cite{barew}.

First we recall how the pomeron part of Eq.~\eqref{BKpom} was derived in~\cite{Bartels:2004ef}.
The linear part on the right hand side of both equations \eqref{BKpom} and \eqref{BKodd} is just the BFKL evolution of the
pomeron and odderon fields.
To see the link  between the reggeized gluon BFKL description and the dipole picture
consider the evolution of the BFKL pomeron wave function in the form of the quasi-Schroedinger equation
\beq
\frac{d}{d Y} \psi(\rho_1,\rho_2) = -H_{12}  \psi(\rho_1,\rho_2)
\eeq
with the kernel $H_{12}=\frac{\bar{\alpha}_s}{2}  h_{12}$, where $\bar{\alpha}_s=\alpha_s N_c/\pi$ and
\begin{equation}
h_{12}=\ln \,\left| p_{1}\right| ^{2}+\ln \,\left| p_{2}\right| ^{2}+\frac{1%
}{p_{1}p_{2}^{\ast }}\ln \,\left| \rho _{12}\right| ^{2}\,p_{1}p_{2}^{\ast
}\,+\frac{1}{p_{1}^{\ast }p_{2}}\ln \,\left| \rho _{12}\right|
^{2}\,p_{1}^{\ast }p_{2}-4\Psi (1)\,.
\label{BFKL_mom}
\end{equation}
Here $\Psi (x)=d\ln \Gamma (x)/dx$,  gluon holomorphic momenta are used and the first two log terms belong to the virtual corrections.
Passing to the coordinate representation and taking into account that the coordinate wave function of the pomeron vanishes when the
two reggeized gluon are located at the same point (termed "the M\"obius representation in ~\cite{Bartels:2004ef,Bartels:2005ji}")
 the action of this kernel on a state can be rewritten as
\begin{equation}
H_{12}\,\psi(\overright{\rho }_{1},\overright{\rho }_{2})=\bar{\alpha}_s\int \frac{d^{2}\rho _{3}}{2\pi }\,\frac{\left| \rho _{12}\right| ^{2}}{%
\left| \rho _{13}\right| ^{2}\left| \rho _{23}\right| ^{2}}\,\left(
\psi(\overright{\rho }_{1},\overright{\rho }_{2})-\psi(\overright{\rho }_{1},\overright{\rho }_{3})-\psi(%
\overright{\rho }_{2},\overright{\rho }_{3})\right)\,,
\end{equation}
This  has the same form as the linear part of Eqs.~\eqref{BKpom} and~\eqref{BKodd}.

Now consider the first non-linear term on the right-hand side of Eq.~\eqref{BKpom}.
The irreducible term $D_4^I$ of Eq.~\eqref{4gluon} contains the transition from $2$ to $4$ reggeized gluon states. The latter evolve in rapidity
with the $4$-gluon Green function, which
in the planar (large $N_c$) limit factorizes into the product of two independent BFKL Green functions: $G_4=G_2 \times G_2$.
Then the second term of Eq.~\eqref{4gluon} acquires the structure of the  triple pomeron vertex.
In the planar limit  $V_{2\to4} D_2$ becomes
\bea
V(1234)D_2 &=&\frac{1}{2}g^2 \Bigl[ G(1,2+3,4)+G(2,1+3,4)+G(1,2+4,3)+G(2,1+4,3)\nonumber\\
&{}&\hspace{-2cm}- G(1+2,3,4)-G(1+2,4,3)-G(1,2,3+4)-G(2,1,3+4)+G(1+2,0,3+4) \Bigr] \nonumber\\
\label{verbart}
\eea
where number $i$ stands from the momentum $q_i$. The general function $G(1,2,3)$ was defined in~\cite{Braun:1997nu}
and consists of two pieces
\begin{equation}
G(\mbox{\boldmath $k$}_1,\mbox{\boldmath $k$}_2,\mbox{\boldmath $k$}_3)=G_1(%
\mbox{\boldmath $k$}_1,\mbox{\boldmath $k$}_2,\mbox{\boldmath $k$}_3)+G_2(%
\mbox{\boldmath $k$}_1,\mbox{\boldmath $k$}_2,\mbox{\boldmath $k$}_3) \, .
\label{Gfunction}
\end{equation}
The first term is constructed from real diagrams
\beq
G_1(\mbox{\boldmath $k$}_1,\mbox{\boldmath $k$}_2,\mbox{\boldmath $k$}_3)=
g^2N_c\,\int \frac{d^2\mbox{\boldmath $q$}_1d^2\mbox{\boldmath $q$}_2}{%
(2\pi)^3} \delta^2 (\mbox{\boldmath $q$}_1+\mbox{\boldmath $q$}_2-\bq) 
W(\mbox{\boldmath $k$}_1,\mbox{\boldmath $k$}_2,\mbox{\boldmath $k$}_3 |\mbox{\boldmath $q$}_1,\mbox{\boldmath $q$}_2)
D_2(\mbox{\boldmath $q$}_1,\mbox{\boldmath $q$}_2),
\eeq
where $\bq=\bk_1+\bk_2+\bk_3$ and
\begin{equation}
W(\mbox{\boldmath $k$}_1,\mbox{\boldmath $k$}_2,\mbox{\boldmath $k$}_3 |\mbox{\boldmath $q$}_1,\mbox{\boldmath $q$}_2)=\!
\left(\!\frac{(\mbox{\boldmath $k$}_2\!+\!\mbox{\boldmath $k$}_3)^2}{(%
\mbox{\boldmath $q$}_1\!-\!\mbox{\boldmath $k$}_1)^2 \mbox{\boldmath $q$}_2^2} \!+\!%
\frac{(\mbox{\boldmath $k$}_1\!+\!\mbox{\boldmath $k$}_2)^2}{\mbox{\boldmath $q$}%
_1^2(\mbox{\boldmath $q$}_2\!-\!\mbox{\boldmath $k$}_3)^2} \!-\!
\frac{ \mbox{\boldmath
$k$}_2^2}{(\mbox{\boldmath $q$}_1\!-\!\mbox{\boldmath $k$}_1)^2 (%
\mbox{\boldmath
$q$}_2\!-\!\mbox{\boldmath $k$}_3)^2} \!-\!\frac{\bq^2}{\mbox{\boldmath $q$}_1^2 %
\mbox{\boldmath $q$}_2^2}\right)  \label{proper}
\end{equation}
The second term is related to the virtual correction present in the
reggeized gluon trajectory:
\bea
G_2(\mbox{\boldmath $k$}_1,\mbox{\boldmath $k$}_2,\mbox{\boldmath $k$}%
_3)=&{}&
- \left[\omega(\mbox{\boldmath $k$}_2)-\omega(\mbox{\boldmath $k$}_2+%
\mbox{\boldmath $k$}_3)\right]
D_2(\mbox{\boldmath $k$}_1,\mbox{\boldmath $k$}_2+\mbox{\boldmath $k$}%
_3)
 \nonumber \\
&{}&
-\left[\omega(\mbox{\boldmath $k$}_2)-\omega(\mbox{\boldmath $k$}_1+%
\mbox{\boldmath $k$}_2)\right] D_2(\mbox{\boldmath $k$}_1+
\mbox{\boldmath $k$}_2,\mbox{\boldmath $k$}_3)\,.\nonumber
\eea
Fourier transforming the vertex~\eqref{verbart},  $D_2$ and the two pomeron subsystems $G_2(12) \times G_2(34)$
(in fact passing to  the M\"obius representation for the states)
one finds
\[
T_4\equiv \langle \psi(12)|(V(1234)D_2) |\psi(34) \rangle
=2g_s^2 \langle \psi(12)|(G(1,2+3,4) |\psi(34)  \rangle\]\beq
=\psi(\rho_{13})\psi(\rho_{23}) \left[ - c
 \frac{\rho_{12}^2}{\rho_{13}^2 \rho_{23}^2}  \right]D_2(\rho_{12}) \,,
 \label{G2pom}
\eeq
where $c=g_s^4 N_c/(4\pi^3)$  and $\rho_{12}=\rho_1\!-\!\rho_2$.
This has exactly the same form of the interaction term present in Eq.~\eqref{BKpom}, with a suitable rescaling (by $\frac{\bar{\alpha}_s} {2\pi c} $ ) of the two
dipole (pomeron) fields $N(\rho_1,\rho_3)N(\rho_2,\rho_3)$.

Now we perform a similar derivation for the last term on the right-hand side, made of two odderon fields.  It comes from the $V_{2\to6}$ vertex present
in the $D_6^I$ of Eq.~\eqref{D6I} discussed before.
Consider the six gluons $(1-6)$ grouped into two odderon fields as $(123)$ and $(456)$. In the planar limit the odderon term of the vertex  can be written in a compact
notation~\cite{bar3} as
\beq
W_6^{odd}(1,2,3|4,5,6)=-\frac{1}{8}g_s^4(\hat{S}_1-\hat{P}_1)
f_{12}(\hat{S}_2^{\dagger}-\hat{P}_2^{\dagger})\,
\label{vertex26}
\eeq
Here the indices 1 and 2 refer to the triplets of gluons (123) and (456),
 the function $f_{12}$ depends on four gluon momenta
\beq
f(1,2|3,4)=G(1,2+3,4)-G(2,1+3,4)-G(1,2+4,3)+G(2,1+4,3) \,
\label{fdef}
\eeq
operator $\hat{S}$ was defined in \eqref{hats} and finally  $\hat{P}$ is an operator which
acting on a function of two gluon momenta antisimmetrizes it in them and
splits the first momentum in three outgoing momenta, while putting to zero the second one:
\beq
\hat{P}(1,2,3|1',2'))\phi(1',2')
= \frac{1}{2}  [\phi(123,0))-\phi(0,123)] \, .
\eeq
(The presence of the subtraction term ($\hat{P}$) in the definition of $W_6^{odd}$  is due to the fact that the function
 $f_{12}$ is not zero in the two 2-gluons subsectors when one the momenta is zero). Calling $\varphi_-$ a generic two gluon antisymmetric function one has ~\cite{blv}
\beq
\langle \Psi | (\hat{S}- \hat{P}) \varphi_{-} \rangle =  c_o   \langle \psi | {\cal N} H_{12} | \varphi_{-} \rangle\,.
\eeq
Note that in the odderon channel the eigenvalues of $H_{12}$ are non negative. The BLV states start with the intercept exactly equal to unity.

One then finds
\beq
\langle\Psi_{n,\nu}(1,2,3) |(\hat{S}-\hat{P})=
 c_o^{1/2} \langle\psi_{n,\nu}(1,2)| H_{12}^{1/2}\, ,
\label{simpli}
\eeq
where $n,\nu$ are the conformal quantum numbers of the odderon state $\Psi$ and the related pomeron state $\psi$.
Using \eqref{simpli}  the contraction of the vertex $W_6^{odd}$ with two odderon states $\Psi_1(1,2,3)$ and $\Psi_2(4,5,6)$
normalized as described above becomes
\[
T_6\equiv -\frac{g_s^2}{8}\langle \Psi_1|(\hat{S}_1- \hat{P}_1)f_{12}(\hat{S}_2^{\dagger}-\hat{P}_2^{\dagger}) |\Psi_2 \rangle\]\[
=-\frac{g_s^2}{8}c_o
\langle \psi_{1\nu_1,n_1} |  H_{12}^{1/2}\, f_{12}  \, H_{12}^{1/2} |\psi_{2\nu_2,n_2} \rangle \nonumber\]\beq
= -\frac{g_s^2}{2}c_o \psi_{\nu_1,n_1}(1,2)  H_{12}^{1/2}\ G(1,2+3,4) H_{34}^{1/2} \psi_{\nu_2,n_2}(3,4)\, ,
\label{Tsimpli}
\eeq
where for the last equality the antisymmetry of the 2 gluon state $\psi$ associated to the odderon state is used.
 Then performing a Fourier transform one finally finds, similarly to Eq.~\eqref{G2pom},
\beq
T_6=\psi(\rho_{13}) H_{13}^{1/2} \psi(\rho_{23}) H_{23}^{1/2} \left[ c'
\frac{\rho_{12}^2}{\rho_{13}^2 \rho_{23}^2} \,\right]  D_2(\rho_{12}),
\eeq
where $c'=\frac{g_s^4 N_c}{16\pi^3}c_o$ and one can identify in the square brackets the pomeron into two odderon vertex in the coordinate representation,
which in Eq.~\eqref{BKpom}  is multiplied by the two odderon fields $O_{\overright{\rho}_1,\overright{\rho}_3}O_{\overright{\rho}_2,\overright{\rho}_3}$.
In order to match to the dipole picture representation in the large $N_c$ limit, as before one needs  to renormalize the BLV odderon field,
rescaling it by the extra operatorial factor $H_{ij}^{1/2}$ as well by the constant factor $\frac{\bar{\alpha}_s}{2\pi\sqrt{c c'}} $.

It is remarkable and important that the pomeron-two-odderons vertex turns out to be of the opposite sign  compared to the triple pomeron vertex.
 In this framework this fact comes automatically from the expressions of the two  vertices $V_{2\to4}$ and $V_{2\to6}$.
This property is related to the different signatures of the pomeron and odderon fields and in particular to the fact that in a pomeron amplitude the two pomeron cut is negative
while the two odderon cut is positive. In the next subsection we show how these properties are realized in an effective local description called Reggeon Field Theory.

In the other Eq.~\eqref{BKodd} the interaction term $ON$ and $NO$ could be computed in the reggeized gluon approach starting from the transition vertex $V_{3\to 5}$,
which unfortunately has not been computed yet in the reggeized gluon approach.

\subsection{A simplified picture: the Reggeon Field Theory}
The equations above are used to study pomeron-odderon fields in the regime where saturation effects set in taming their growth with rapidity.
This behavior is induced by the fan structure which is resummed by the differential equations.
Nevertheless, as always stressed by L.N.~Lipatov, the real picture is more complicated since one should include also loops involving
pomeron and odderon fields (and more in general reggeized gluons).
Moreover at high energy (large rapidities) but long transverse distances non perturbative QCD physics, where hadrons are involved, cannot be ignored.
This is a problem whose solution is well out of reach at present state.

Before the QCD era strong interactions were studied in the context of S-matrix and the scattering in Regge limit was described by simpler systems derived from the analytic structure of the partial wave amplitudes, in particular focusing on the so called reggeons associated to their poles. As the next natural step, reggeon interactions were considered by V.N.~Gribov and encoded in a $2+1$ dimensional Reggeon Field Theory, with one time (rapidity) and two transverse space dimensions.
This is a framework were pomeron and odderon interactions can be studied, in particular using renormalization group (RG) techniques.
Recently an investigation using a functional RG approach was carried on~\cite{Bartels:2015gou,  Bartels:2016ecw},
including the case of a system of one pomeron and one odderon local fields, denoted as $\psi,\psidag$ and $\chi,\chidag$ respectively.
The RFT action in its simpler version, with just ultra local cubic interactions, has the form:
\bea
S[\psidag,\psi,\chidag,\chi]&=&\int \!  \, \mathrm{d}^D x \,  \mathrm{d} \tau
 \Bigl( Z_{P }(\frac{1}{2} \psi^{\dag} \overset\leftrightarrow{\partial}_{\! \tau} \psi -\alpha'_{P } \psidag\nabla^{2}\psi)
+ Z_{O }(\frac{1}{2}\chidag\overset\leftrightarrow{\partial}_{\! \tau}\chi -\alpha'_{O } \chidag\nabla^{2}\chi)\nonumber\\
&{}&+ V[\psi,\psidag,\chi,\chidag] \Bigr)\,.
\eea
with the potential $V=V_3$ given by
\beq
V_3=-\mu_P \psidag \psi +i\lambda \psidag (\psi +\psidag) \psi
  -\mu_O \chidag \chi +i\lambda_2 \chidag (\psi +\psidag) \chi + \lambda_3
(\psidag \chi^2 + {\chidag}^2 \psi).
\label{uptocubic}
\eeq

Here $D$ is the number of spatial dimension ($D=2$ is the physical case), which can be conveniently considered as a continuous parameter.
The potential (allowed interactions) is constrained by signature conservation (even for the Pomeron and odd for the Odderon) and by the overall sign of the multi-Reggeon discontinuity amplitudes $-i \prod_j (i \xi_j)$, where $\xi_j$ are the signature factors $\xi=(\tau-e^{-i\pi \omega})/\sin{\pi\omega}$ with $\omega=\alpha(0)-1$: note that for the Pomeron $\xi_P$ is almost imaginary while for the Odderon $\xi_O$ is almost real.
This implies that $t$-channel states with odd and even number of Odderons never mix. Another constraint on the potential is that transitions $P\to PP$ are imaginary (two Pomeron cut is negative), $P\to O O$ are real (two Odderon cut is positive) and  $O\to O P$ is imaginary (Odderon-Pomeron cut is negative). These considerations, implemented in Eq.~\eqref{uptocubic}, are easily generalized to higher order (subleading) interactions~\cite{Bartels:2016ecw}.
As a consequence we can write the potential in terms of different contributions with operators which relate states differing by an integer number of Odderon pairs.
In perturbative QCD the $P\to O O$ vertex has been computed in the generalized leading-log approximation. We shall see that the Reggeon interaction nevertheless are dominated by a fixed point in which operators changing the number of Odderon pairs are not present, these being present only in deformations of the critical theory.

Since $D=4$ is the scaling dimension (critical dimension) of reggeon field theory, critical behaviour can be studied in perturbation theory
with the $\epsilon$ expansion in $D=4-\epsilon$ dimensions. Moreover functional renormalization group techniques based on the use of the effective average action can be used in $D=2$ in a truly non perturbative context.
We present here the results of a one-loop renormalization group analysis in $D=4-\epsilon$ around the critical dimension $4$ of the transverse space. We do not report the beta functions here but just the results for the fixed points and the linear behavior around them. In~\cite{Bartels:2016ecw} also the case $D=2$ was studied using non perturbative methods based on functional renormalization group techniques, which shows that the perturbative results are qualitatively correct but quantitatively different.
Using the cubic truncation one finds, besides a fixed point solution related to the pure Pomeron theory, a second non trivial fixed point also in the Odderon sector, such that $\lambda^2, \lambda_2^2,  \lambda_3^2, \mu_P, \mu_O=O(\epsilon)$:
\bea
\label{specialFPepsilon}
&{}&\!\!\!\!\mu_P=\frac{\epsilon}{12},\quad \mu_O=\frac{95\!+\!17\sqrt{33}}{2304}\epsilon, \quad \lambda^2=\frac{8\pi^2}{3} \epsilon, \quad \eta_P=-\frac{\epsilon}{6}, \quad  \eta_O=-\frac{7\!+\!\sqrt{33}}{72}\epsilon,\nonumber\\
&{}&\!\!\!\! \lambda_2^2=\frac{23\sqrt{6}\!+\!11\sqrt{22}}{48}\epsilon, \quad \lambda_3=0,
 \quad  \zeta_P=\zeta_O=\frac{\epsilon}{12},
\quad r=\frac{3}{16}(\sqrt{33}\!-\!1).
\eea
Here $\eta_{P,O}=-\mu\frac{d}{d \mu} \log Z_{P,O}$ are the corresponding field anomalous dimensions,
$\zeta_{P,O}=-\mu\frac{d}{d \mu} \log \alpha'_{P,O}$ are the corresponding anisotropic anomalous dimensions and
$r=\alpha'_O/\alpha'_P$ is the ratio between the slopes.
One immediately notes that $\lambda_3=0$ so that the critical theory has no operator changing the number of Odderon pairs.
Apart from the anomalous dimensions, the universal critical exponents are extracted with a spectral analysis of the stability matrix,
obtained perturbing the beta functions around the fixed point.
We find two negative eigenvalues, associated to two relevant directions, which give the standard corresponding critical exponents $\nu$ related to the behavior of the two point functions for the pomeron and odderon:
\bea
\lambda^{(1)}=-2+\frac{\epsilon}{4} \,\,\,\rightarrow \nu_P=\frac{1}{2} +\frac{\epsilon}{16} \quad , \quad
\lambda^{(2)}=-2+\frac{\epsilon}{12} \rightarrow \nu_O=\frac{1}{2} +\frac{\epsilon}{48}.
\label{nuepsilon}
\eea
This fixed point (as well as its non perturbative counterpart for $D=2$) is important in order to understand the renormalization group flow properties.
Indeed we expect that from the QCD description at short distance moving to larger and larger transverse distances one gets closer to an effective RFT formulation.
As a last remark we can add that this system is in the same universality class of an out of equilibrium directed percolating system,
as was shown to be the case by J. Cardy for a purely single pomeron RFT.

\section{The running coupling constant}
\subsection{Lipatov's proposal}
The perturbative QCD pomeron and odderon in the LLA come from a summation of terms of the order
$(\alpha_s\ln s)^n$ but neglect terms of the order $(\alpha_s\ln k^2)^n$ responsible for
the running of the coupling. So in this theory $\alpha_s$ is assumed to be a fixed (small) parameter.
As a result the BFKL theory neglects one of the basic properties of the QCD - the running of the coupling and so the
asymptotic freedom. It also automatically neglects its other basic properties - the confinement, which manifests itself
in the growth of the coupling at small $k^2$. It is then natural that nearly immediately after the introduction of the
BFKL pomeron attempts to introduce the running of the coupling in the theory appeared .

The first of these attempts was made by L.N.~Lipatov as early as in 1986 ~\cite{lip1}. It was rather straightforward.
Let the BFKL Hamiltonian $H$ be written with the separated factor $\alpha_s$, namely
$H=\alpha_sh$. Consider the forward case, so that the BFKL equation is given by
\beq
\alpha_s(h\psi)(k)=E\psi(k),\ \  {\rm or}\ \ (h\psi)(k)=\chi\,\psi(k)
\label{liprun0}
\eeq
with the set of eigenvalues $\chi_{\nu,n}$ and eigenfunctions  $\psi_{\nu,n}$
\beq
\chi_{\nu,n}=2\hat{\psi}(1)-\hat{\psi}\Big(\frac{1}{2}+i\nu\Big)-\hat{\psi}\Big(\frac{1}{2}-i\nu\Big),
\ \ \psi_{\nu,n}(k)=ck^{2i\nu-1/2}e^{in\phi} \,,
\eeq
where $\hat{\psi}$ is the logarithmic derivative of the $\Gamma$-function,
and $-\infty<\nu<+\infty$, $n=0,\pm 1,\pm 2,...$.

The idea of L.N.~Lipatov was to change the fixed $\alpha_s$ in (\ref{liprun0}) into the running one as a function of $k$
\beq
\alpha_s \to\frac{1}{\beta_0 \ln(k^2/\Lambda_{QCD}^2)},\ \ \beta_0=\frac{11N_c-2N_f}{12\pi}=\frac{b}{\pi} \,.
\label{liprun}
\eeq
Rescaling $k\to k/\Lambda_{QCD}$ Eq. (\ref{liprun0}) then turns into
\beq
(h\psi)(k)=\omega \, \beta_0 \ln(k^2) \psi(k).
\label{lipeq1}
\eeq
Now one passes to the $(\nu,n)$ representation
\beq
\psi(k)=\int d\nu\sum_n\psi_{\nu,n}a_{\nu,n}
\eeq
and obtains an equation for $a_{n,nu}$
\beq
i\omega \beta_0 \frac{d a_{\nu,n}}{d\nu}=\chi(\nu,n)a_{\nu,n}
\label{lipeq2}
\eeq
which evidently admits an explicit solution.

The next step is to impose the appropriate boundary conditions. Addressing the reader to the original paper for details, we only
mention that they are two, following from the behavior of the resulting $\psi(k)$ at large and small $k$.
At large $k$ it is required that the solution
pass to the known form for the DGLAP equation in the double log approximation. At small $k$ it is required that the phase of $\psi$
should be fixed from the non-perturbative region. The two boundary conditions make the spectrum a discrete one, which for
the $n$-th state with $n>>1$ is
\beq
\omega_n=-\frac{0.4085}{n-\frac{1}{4}+\frac{\phi_0-\nu t_0}{\pi}}\,,
\label{lipeq3}
\eeq
where $\phi_0$ is the non-perturbative phase and  small $t_0=k_0^2/\Lambda_{QCD}^2$ is the point at which the low boundary
condition is imposed.
As a result one finds an infinite sequence of poles in the $j$ plane converging to the point $j=1$.

This way of introduction of the running coupling raises questions. Operator $h$ is in fact an integral one with a kernel $h(k,k')$.
Change (\ref{liprun}) implies that one  argument is used for the running. Such a choice, apart from being arbitrary, violates the
intrinsic symmetry between $k$ and $k'$ essential for the interchange projectile $\lra$ target. The form of the boundary
at small $k$ does not seem to be unique. Still the conclusion of the splitting of the cut into a series of poles as a result of the
running of the coupling was later confirmed in other versions of the running coupling and so was an important prediction
in this line of approach. Also the Lipatov running coupling allows to obtain many explicit results, in contrast to more
elaborate versions.
Later it was extensively  used to obtain physically important results (the latest are~\cite{kowalski1, kowalski2, kowalski3}).
See also~\cite{Bartels:2018pin}.

\subsection{The running coupling constant from the bootstrap}\label{sec:two_gluons}
In search of a more fundamental basis for the introduction of the running coupling constant many years ago we turned to the bootstrap
property of the BFKL dynamics, which is one of the pillars on which the idea of gluon reggeization rests ~\cite{brarun}. As is well known, the bootstrap
discovered in the original papers ~\cite{bfkl1,bfkl2} is the unitarity condition for the channel with the gluon color number. It guarantees
that the unitarity in this section gives rise to the solution in the form of the reggeized gluon. It was natural to expect that
this condition should be preserved also with the running coupling. So preservation of the bootstrap could be the guiding principle for the
introduction of the running coupling constant. Later the idea that a strong bootstrap condition in perturbation theory up to next-to-leading order was propsed in~\cite{Braun:1998zj, Braun:1999uz},
where non trivial relations for the NLO BFKL in the octet channel, the reggeized gluon trajectory and any inpact factor were conjectured. They were later on veryfied by V.~Fadin and collaborators~\cite{Fadin:2002hz}.

To realize this idea one can write the inhomogeneous equation for two reggeons as
\beq
(H-E)\psi =F,
\label {scheq}
\eeq
with the Hamiltonian for the color group $SU(N_{c})$
\beq
H=-\omega(q_{1})-\omega(q_{2})+(T_{1}T_{2})V.
\label{ham2}
\eeq
The form of $\omega(q)$ and interaction $V$ are in fact interrelated. To make it explicit we
express both via a single function $\eta(q)$.\footnote{We stress that this is just a possible way, but not the unique one, to satisfy the bootstrap relation.}
Namely
\beq
\omega(q)=\frac{1}{2}N_c\int \frac{d^2q_1}{(2\pi)^2}\frac{\eta(q)}
{\eta(q_1)\eta(q-q_1)}.
\label{traj}
\eeq
 The interaction term, apart from the product of the
gluon color vectors $T_{i}^{a},\ i=1,2,\ a=1,...N_{c}$, involves
the interaction kernel $V$ which we also express via the same
function $\eta(q)$
\beq
V(q_1,q_2|q'_1q'_2)=\frac{\eta(q_1+q_2)}{\eta(q'_1)\eta(q'_2)}
-\Big(\frac{\eta(q_1)}{\eta(q'_1)}+
\frac{\eta(q_2)}{\eta(q'_2)}\Big)\frac{1}{\eta(q_1-q'_1)}.
\label{int}
\eeq
It conserves the momentum: $q_1+q_2=q'_1+q'_2$.
Combining (\ref{traj}) with (\ref{int}) for the gluon channel with $(T_1T_2)=-\frac{1}{2}N_c$ one finds the bootstrap
relation for any $\eta(q)$:
\beq
\frac{1}{2}N_c\int\frac{d^{2}q_{1}}{(2\pi)^{2}}V(q_1,q_{2}|q'_1,q'_2)=\omega(q_1)+\omega(q_2)-\omega(q_{1}+q_{2})\,.
\label{boots}
\eeq
Then taking in the inhomogeneous equation $F=F(q_1+q_2)$ one finds a solution to
 Eq. (\ref{scheq}):
\beq
\psi(q_1,q_2)=\psi(q_1+q_2)=\frac{F(q_1+q_2)}{\omega(q_1+q_2)-E}\,.
\eeq
This means that the two gluons 1 and 2 have fused into a single one with
the momentum $q_1+q_2$ and the moving pole at $j=1-E$. This phenomenon is true for arbitrary $\eta(q)$.

For the original BFKL equation with the fixed coupling we have a particular $\eta(q)$:
\beq
\eta(q)=\frac{2\pi}{g^2}q^2, \ \ \eta(0)=0 \,.
\label{etafix}
\eeq
Then one finds the standard LL BFKL expressions for the trajectory $\omega(q)$
and interaction $V(q_1,q_2|q'_1,q'_2)$. Also the trajectory passes zero at $q=0$ to correspond to  the real gluon with
a vanishing mass.

Using the fact that the bootstrap is fulfilled for arbitrary $\eta(q)$ we can choose $\eta(q)$
to satisfy the high-momentum behavior of the gluon distribution with a running
coupling~\cite{brarun}
\beq
\eta(q)=\frac{b}{2\pi}\, q^2\ln\frac{q^2}{\Lambda_{QCD}^2},\ \ q^2>>\Lambda^2,
\label{asym}
\eeq
where $\Lambda$ is the standard QCD parameter and $b=(11Nc-2Nf)/12$ is defined in (\ref{liprun}).

The asymptotic (\ref{asym}) and condition $\eta(0)=0$ are the only
properties of $\eta(q)$ which follow from the theoretical reasoning.
A concrete form of $\eta(q)$ interpolating between $q^2=0$ and
(\ref{asym}) may be chosen differently. One hopes that
physical results will not too strongly depend on the choice.

With $\eta(q)$ satisfying (\ref{asym}) we can build the trajectory $\omega(q)$ and the interaction
$V$ between two reggeons with the running coupling constant. In this way we obtain equations for both pomeron,
odderon and higher BKP states with the running coupling constant. Note that as with the fixed coupling the
resulting equations remain infrared stable if $\eta(0)=0$ as dictated by the zero gluon mass. However we can
improve the infrared behavior to mimic confinement and introduce a finite gluon mass $m$ dropping the latter condition.
 In our practical applications we use
\beq
\eta(q)=\frac{b}{2\pi} f(q),\ \ \ f(q)=(q^2+m^2)\ln\frac{q^2+m^2}{\Lambda^2} \,.
\label{eta1}
\eeq
With this choice $\eta(0)\neq 0$ and has to be taken into account in the interaction.

It is interesting to note that this very simple ansatz given in Eqs.~\eqref{traj} and \eqref{int} in order to satisfy the bootstrap condition,
when feeded with the choice of Eq.~\eqref{asym} can be studied to compare to the results of the NLL BFKL kernel and the external particle impact factors.
This can be conveniently done using dimensional regularization~\cite{Braun:1999uz} where it was shown that
fermionic contributions (the $N_f$ dependent part) was correctly reproduced, as well as the gluon sector related to the running coupling. Only a gluon contribution of the NLL approximation is actually missing.
This is therefore an interesting approximation containing an important part of the NLL corrections and can be considered
as a good starting point to study the running coupling effects.

\subsection{The BLV odderon with the running coupling constant}\label{sec:three_gluons}

For the odderon
the  color wave function is $d_{a_1a_2a_3}$ where $a_i$ is the
color of the $i$-th reggeon. Separating it, the  equation for the odderon momentum wave function is
\beq
(H_{od}-E)\psi_{od}=0,
\label{odeq}
\eeq
with a Hamiltonian
\beq
H_{od}=-\sum_{i=1}^{3}\omega(q_i)-\frac{1}{2}N_c\sum_{i<k}^{3}V_{ik}.
\label{hamod}
\eeq

 Consider the odderon
equation with a specific inhomogeneous term
\beq
(H_{od}-E)\psi_{12}=F_{12},
\label{odinho}
\eeq
where
\beq
F_{12}(1,2,3)=\int \frac{d^{2}q'_{3}}{(2\pi)^{2}}
\hat{W}(1,2,3|1',3'))\psi(1',3')
+F_{2}(1,2,3),
\label{inho}
\eeq
where for brevity we denote momenta $q_1,q_2,q_3$ just by their numbers 1,2,3 and so on.
Here $q_1'+q_3'=q_{1}+q_{2}+q_{3}$. In Eq. (\ref{inho})
$\psi (1',3')$ is some function of two momenta $q'_1$ and $q'_3$.
The kernel  $\hat{W}$ is a sum of two terms
\beq
\hat{W}(1,2,3|1',3')=-\frac{1}{2}N_c
\Big(W(2,1,3|1',3')+W(1,2,3|1',3')\Big)
\label{opw}
\eeq
and $W$ is a difference between two kernels
(\ref{int}), expressed via function $\eta(q)$.
In full notation
\beq
W(q_{2},q_{1},q_{3}|q'_{1},q'_{3})=
V(q_{1},q_{3}|q'_{1}-q_{2},q'_{3})
-V(q_{1}+q_{2},q_{3}|q'_{1},q'_{i}).
\label{opw1}
\eeq
 It possesses
an important  symmetry property:
\beq
W(1,2,3|1',3')=W(3,2,1|3',1').
\label{symw}
\eeq
Note that in the fixed coupling case this kernel reduces to
the Bartels kernel $K_{2\to 3}$ which describes transition from two
BFKL pomerons to three ~\cite{barew}. Therefore one could be tempted to introduce in a similar way the running coupling also in the effective vertices which change the number of ($t$-channel) reggeized gluons,
and in particular the pomeron $\to$ 2 pomeron and the pomeron $\to$ 2 odderon vertices whose expressions are known at fixed coupling~\cite{barwue,barew}.

It can be shown ~\cite{brarun1} that the inhomogeneous equation  (\ref{odinho})
can be solved by
\beq
\psi_{12}(q_1,q_2,q_3)=\psi(q_1+q_2,q_3)
\eeq
provided function $\psi$ is a pomeron wave function
which satisfies the homogeneous Schroedinger equation (\ref{scheq})
with a Hamiltonian   (\ref{ham2}) in which $T_1T_2=-N_c$ (the BFKL equation
in the case of a fixed coupling).

Using this,  we cyclically permute the gluons 1,2,3 to obtain two more relations
\beq
(H_{od}-E)\psi_{23}=F_{23}
\label{odeq23}
\eeq
and
\beq
(H_{od}-E)\psi_{31}=F_{31},
\label{odeq31}
\eeq
where for instance
\beq
\psi_{23}(q_1,q_2,q_3)=\psi(q_2+q_3,q_1)
\eeq
and
\beq
F_{23}(1,2,3)=-\frac{1}{2}N_c\int \frac{d^{2}q'_{3}}{(2\pi)^{2}}
\Big(W(3,2,1|1',3')+W(2,3,1|1',3')\Big)\psi(1',3')
\label{inho23}
\eeq
(the integration momenta $q'_3$ and $q'_1=q_1+q_2+q_3-q'_3$ do not change
under permutations of the external momenta).
We add all the three  equations obtained in this manner
together  to obtain
\beq
(H_{od}-E)\Big(\psi_{12}+\psi_{23}+\psi_{31}\Big)=
F_{12}+F_{23}+F_{31} \equiv F_{tot}.
\label{odeqtot}
\eeq
The total inhomogeneous term is
\beq
F_{tot}(1,2,3)=-\frac{N_c}{2}\int\frac {d^{2}q'_{3}}{(2\pi)^{2}}
U(1,2,3|1',3')\psi(1',3')
\label{inhotot}
\eeq
where
\[
U(1,2,3|1',3')=
W(2,1,3|1',3')+W(1,2,3|1',3')+W(3,2,1|1',3')\]\beq
+W(2,3,1|1',3')+
W(1,3,2|1',3')+W(3,1,2|1',3').
\eeq
Using property (\ref{symw}) one easily finds that $U$ is symmetric in the
last pair of arguments
\beq
U(1,2,3|1',3')=U(1,2,3|3',1').
\eeq
It follows that if function $\psi(q_1,q_2)$ is antisymmetric in its
variables, the total inhomogeneous term  (\ref{inhotot}) in Eq.
(\ref{odeqtot}) vanishes and the sum
\beq
\psi_{od}=\psi_{12}+\psi_{23}+\psi_{31}
\label{odsol}
\eeq
is a solution of the odderon equation
\beq
(H_{od}-E)\psi_{od}=0.
\label{eqod}
\eeq
Thus from any antisymmetric solution of the pomeron equation one gets
a corresponding solution of the odderon equation.

For a fixed coupling this result was found in ~\cite{blv}. Our derivation
generalizes it to the case of a running coupling, provided it is
introduced in the manner which preserves the bootstrap.

\subsection{A general family of BKP states in the planar limit with the running coupling constant}\label{sec:bkp}
We consider here an $n$ gluon composite state in color singlet which has the high energy behavior in the regge limit described by the following kernel
\beq
H_n = - \sum_{i=1}^n \, \omega(q_i)+ \sum_{i<j} T_i T_j \, V_{ij}\,
\label{bkp_ham}
\eeq
which in the planar limit reduces to
\beq
H_n^{\infty} = \frac{1}{2} \left[ H_{12}^{(1)} + H_{23}^{(1)}
+ \cdots + H_{n1}^{(1)} \right]
\label{bkpnlarge}
\eeq
and where $H_{ij}$ is a two gluon Hamiltonian acting on the gluons $i$ and $j$.
As for the case of the previous section we shall proceed with the replacement from the fixed coupling case of Eq.~\eqref{etafix} to the running coupling one of Eq.~\eqref{asym},
this time following an approach similar to~\cite{Vacca:2000bk}.
Starting from the ansatz
\beq
\psi_n(k_1,k_2,\cdots,k_n)=
\sum_{i=0}^{n-1}  (R_n)^i c_i \, \psi_{n-1}(k_1+k_2,k_3,\cdots,k_n),
\label{ampuans}
\eeq
where $R_n$ is the generator of a cyclic shift along the cylinder, i.e. $(R_n)^n=1$ and $c_i$ are constant to be determined,
and rearranging the Hamiltonian in the form
\beq
\bar{H}_n^\infty(1,2,\cdots,n)  =
\frac{1}{2} \left( \bar{H}_{12}^{(1)} + \bar{H}_{1n}^{(1)}
- \bar{H}_{2n}^{(1)} \right) + \bar{H}_{n-1}^\infty(2,3,\cdots,n),
\label{decomp}
\eeq
let us consider the action of the Hamiltonian on the function $g = \psi_{n-1}(k_1+k_2,k_3,\cdots,k_n)$. One finds
\bea
&&\bar{H}_n^\infty
(1,2,\cdots,n) \otimes g =
\nonumber \\
&&\int \{d^2 k_i'\} \bar{H}_{n-1}^\infty
(k_1+k_2,k_3,\cdots,k_n|k_2',k_3',\cdots,k_n')
\, \psi_{n-1}(k_2',k_3',\cdots,k_n') + \nonumber \\
&& \frac{1}{2} \int \{d^2 k_i'\}\,  W(k_1,k_2,k_3|k_1',k_3')
\psi_{n-1}(k_1',k_3',\cdots,k_n') + \nonumber \\
&&\frac{1}{2} \int \{d^2 k_i'\}\,  W(k_2,k_1,k_n|k_2',k_n') \,
\psi_{n-1}(k_2',k_3',\cdots,k_n'),,
\label{collect}
\eea
where we denote with $\{d^2 k_i'\}$ the necessary integral measure. Imposing for $\psi_{n-1}$ the following properties:
(1) it is an eigenstate, i.e. $\bar{K}_{n-1}^\infty \, \psi_{n-1}= E \, \psi_{n-1}$ and
(2) has definite symmetry such that $R_{n-1} \, \psi_{n-1}= r_{n-1} \, \psi_{n-1}$,
one can write the previous relation as
\bea
&&\bar{H}_n^\infty
(1,2,\cdots,n) \otimes \psi_{n-1}(k_1+k_2,k_3,\cdots,k_n) =
E \, \psi_{n-1}(k_1+k_2,k_3,\cdots,k_n) + \nonumber \\
&& \frac{1+r_{n-1} (R_n)^{-1}}{2} \int \{d^2 k_i'\}\,
W(k_1,k_2,k_3|k_1',k_3') \psi_{n-1}(k_1',k_3',\cdots,k_n')
\label{compact}
\eea
so that, using the definition of Eq.~\eqref{ampuans} we can have the relation
\beq
\bar{H}_n^\infty   \psi_n= E \, \psi_n
\eeq
provided the unknown coefficients $c_i$ satisfy the secular equation
\beq
\sum_{i=0}^{n-1}  (R_n)^i \Bigl( 1 +r_{n-1} (R_n)^{-1} \Bigr) \, c_i =0\,,
\eeq
where $r_{n-1}$ plays the role of eigenvalue and $c_i$ are the components of the corresponding eigenvector.
Using the constraint $(r_{n-1})^{n-1}=1$, one finds $r_{n-1}=(-1)^n$.
Therefore for $n$ even one has $r_{n-1}=+1$ and $c_i=(-1)^i$ while for $n$ odd we get $r_{n-1}=+1$ and $c_i=+1$.
The case with $n$ even is unphysical because the state $\psi_n$ would be not Bose symmetric.
Instead the case with $n$ odd is physical, with the solution constructed from another solution for $n-1$ gluons which is odd under cyclic permutations.
In particular the odderon corresponds to the special case $n=3$.
In summary, given for an even number of gluons a solution of the kernel, improved to include running coupling effects using the bootstrap approach,
with some intercept eigenvalue, then one can automatically construct another solution with one more gluon with the same intercept.

\subsection{Equations for the pomeron and BLV odderon with the running coupling}
The BFKL equation admits explicit solutions both in the forward and non-forward
directions based on its conformal invariance. With the running coupling this invariance is lost
and one has to relay on numerical studies. The first calculations of the pomeron ground states
intercept and slope were done nearly immediately after the introduction of the running coupling from the bootstrap
~\cite{bvv}. Here we expand these calculations and extend them to the BLV odderon. We borrow from ~\cite{bvv}
the starting equations in the form suitable for numerical calculations.

We put  $N_c=3$,  present $\eta(q)$  according to (\ref{eta1}) and remove all numerical factors including $b$ from the Hamiltonian
redefining the eigenvalue. We also
symmetrize the integral kernel in $V$ by introducing a new wave function $\phi$ by
\beq
\psi(q_1)=\phi(q_1)\sqrt{f(q_1)f(q_2)},\ \  q_1+q_2=q
\eeq
where $q$ is the total momentum.
Then the  equation for $\phi$ takes the
form
\beq
\int d^{2}q'_{1} h_q(q_1,q'_1) \phi(q'_{1})= A_{q}(q_{1})\phi(q_{1})+\int d^{2}q'_{1}L_{q}(q_{1},q'_{1})\phi(q'_{1})=
\epsilon(q)\phi(q_{1})
\label{basic}
\eeq
Here the "kinetic energy" is
\beq
A_{q}(q_{1})=(1/2) \int\frac{d^{2}q'_{1}\fa}{\fc f(q_{1}-q'_{1})}+
(1/2)\int\frac{d^{2}q'_{2}\fb}{\fd f(q_{2}-q'_{2})}\,.
\label{a0q}
\eeq
The interaction kernel consists of two parts, a quasilocal  and a
separable ones:
\beq
L=L^{(ql)}+L^{(sep)}
\label{lq}
\eeq
They are given by
\beq
L^{(ql)}_{q}(q_{1},q'_{1})=
-\sqrt{\frac{\fa}{\fb}}\frac{1}{f(q_{1}-q'_{1})}\sqrt{\frac{\fd}{\fc}}-
\sqrt{\frac{\fb}{\fa}}\frac{1}{f(q_{2}-q'_{2})}\sqrt{\frac{\fc}{\fd}})
\label{lql}
\eeq
and
\beq
L^{(sep)}_{q}(q_{1},q'_{1})=\frac{\fq}{\sqrt{\fa\fb\fc\fd}}\,.
\label{lsep}
\eeq
Both parts are evidently symmetric in $q_{1}$ and $q'_{1}$.
The scaled energy $\epsilon$ is related to the initial one by
\beq
E=\frac{3}{2\pi b} \epsilon= \frac{6}{\pi (11-(2/3)N_{F})}\epsilon\,.
\label{energy}
\eeq

Solutions of Eq. (\ref{basic}) give the pomeron energies and wave function. Its solutions
antisymmetric in $q_1\lra q_2$ give energies and wave functions of the BLV odderon.

The nonforward equation (\ref{basic}) with $q\neq 0$ is too complicated for the numerical study
due to three independent variables. So we concentrate on the forward case $q=0$.

At $q=0$ the
equation retains its form (\ref{basic}) with
\beq
A_{0}(q_{1})=\int\frac{d^{2}q'_{1}\fa}{\fc f(q_{1}-q'_{1})}
\eeq
and the interaction given by (17) where now
\beq
L^{(ql)}_{0}(q_{1},q'_{1})=
-\frac{2}{f(q_{1}-q'_{1})}
\eeq
has really become local and
\beq
L^{(sep)}_{0}(q_{1},q'_{1})=\frac{f(0)}{\fa\fc}\,.
\eeq
In the following we omit the subindex 0 implying $q=0$.

To keep under control the behavior of the kernel at extremely large values of momenta we transform the
trajectory term $A(q)$ similar to what is standardly done in the BFKL equation.
Using the identity
\[\int d^2q_{1} \frac{\fq}{\fa f(q-q_1)}=2\int d^2q_{1} \frac{\fq}{f(q-q_1)\Big(\fa+ f(q-q_1)\Big)},\]
we find
\beq
A(q)=\int dq_{1}^{2}\int_0^{2\pi}d\chi_1 \frac{\fq}{f(q-q_1)\Big(\fa+f(q-q_1)\Big)}.
\label{a02}
\eeq
where $\chi_1$ is the azimuthal angle of $q_1$.

To make the equation for $\phi$ one-dimensional, the angular momentum $n$ of the gluons is introduced
\beq
\phi(q)=\phi_{n}(q^{2})e^{i n\chi}\,,
\eeq
where $\chi$ is the azimuthal angle. Integrating over it in the
equation, one obtains an one-dimensional integral equation for the radial
function $\phi_{n}(q^{2})$:
\beq
A(q)\phi_{n}(q^{2})+\int dq_{1}^{2}
 L_{n}(q^{2},q_{1}^{2})\phi_n(q_{1}^{2})=
\epsilon\phi_{n}(q^{2})\,,
\label{eq1}
\eeq
with the kernel $L_n$  now given by
\beq
L_{n}(q^{2},q_{1}^{2})=-B_{n}(q^{2},q_{1}^{2})
+\delta_{n0}\pi\frac{f(0)}{\fq\fa}\,,
\label{ln}
\eeq
where
\beq
B_{n}(q^{2},q_{1}^{2})=\int_{0}^{2\pi} d\chi
\frac{\cos n\chi}{f(q^{2}+q_{1}^{2}-2qq_{1}\cos\chi)}\,.
\label{bn}
\eeq
The kinetic term $A(q)$  after the angular integration in Eq.~(\ref{a02}) can be written as
\beq
A(q)=\int dq_{1}^{2}\frac{\fq}{\fa}\Big(B_0(q,q_1)-C_0(q,q_1)\Big),
\label{a03}
\eeq
where
\beq
C_0(q,q_1)=\int_0^{2\pi}d\chi\frac{1}{\fa+f(q-q_1)}.
\label{c0}
\eeq

These are the equations we are solving numerically. We shall see that the ground state of the pomeron
with negative energies indeed splits into an infinite  series of discrete states, with energies
converging to $E=0$. In contrast the odderon states remain continuous with energies along a cut in the
$j$ plane starting from $j=1$. As a result we expect that in the non-forward direction the discrete
pomeron states will move with the total momentum $q$ with the intercept $\alpha(q)$.
As to the continuous odderon states we expect them to stay on the cut above $E=0$. Should they move, it
may introduce moving unphysical cuts in the $t$-plane and violate the unitarity.

As mentioned the equations become very difficult to solve at $q\neq 0$ even numerically.
So, rather than to attempt to solve them for all $q$, we limit ourselves
to small values of $q$ and determine not the whole trajectory $\alpha(q)$
but only the intercept $\alpha'$ defined by
\beq
\alpha(q)=1-E(q)=1+\Delta-\alpha'(0)q^2.
\eeq
This can be done in a much
simpler manner using a perturbative approach. We present "the Hamiltonian"
in (\ref{basic})
\beq
H_{q}=A_{q}+L_{q}=H_{0}+W(q)
\eeq
and calculate analytically $W(q)$ up to terms of the second order in $q$.
Then for small $q$ the value of the energy $\epsilon(q)$ will be given
 by the standard perturbation formula
\beq
\epsilon(q)=\epsilon (0)+\langle W(q) \rangle\,,
\label{epqu}
\eeq
where $\langle \rangle$ here means taking the average with the wave function at $q=0$,
determined from the numerical solution of the equation discussed above.
Note that for the continuous and non-normalizable states $\langle W(q) \rangle=0$
is expected  unless the interaction is too bad. Our numerical studies confirm this
result.

Technically it is natural to write the momenta as
\beq
q_1=\frac{1}{2}q+l, \quad q_2=\frac{1}{2}q-l, \quad q'_1=\frac{1}{2}q+l',\quad q'_2=\frac{1}{2}q-l'
\label{qll}
\eeq
and expand in power series all expressions in (\ref{basic}) up to the second order in $q$. The corresponding calculations are
straightforward but tedious and so together with the result they are briefly presented in the Appendix.

\subsection{Change of variable and discretization}
We pass to the logarithmic variable
$t=\ln q^2$ (the unit in which $q^2$ is measured is inferred from the numerical values of the dimensional parameters $m$ and $\Lambda_{QCD}$).
To retain the symmetry of the kernel we express $\psi(q)$ as
\beq
\phi_n(q^2)=\tilde{\phi}_n(q^2)/q.
\eeq
The equation becomes
\beq
A(q)\tilde{\phi}_{n}(q^{2})+\int dt_1 q q_1
 L_{n}(q^{2},q_{1}^{2})\tilde{\phi}_n(q_{1}^{2})=
\epsilon\tilde{\phi}_{n}(q^{2}),
\label{eq2}
\eeq
where $q^2=e^t$ and $q_1^2=e^t_1$.
Here
\beq
A(q)=
\int dt_1q_{1}^{2}\frac{f(q)}{\fa}\Big(B_0(q^2,q_1^2)-C_0(q^2,q_1^2)\Big).
\label{a04}
\eeq

The discretization is realized using a uniform grid in $t$
\beq
t_i=t_{min}+id,\ \ i=0,1,\cdots   , N,\quad d=\frac{t_{max}-t_{min}}{N}.
\label{ti}
\eeq
We approximate the integrals over $t$ by finite sums
\beq
\int_{-\infty}^{\infty}dt\,F(t)\simeq\sum_{i=0}^{N}w_{i}F(t_{i})
\label{apprint}
\eeq
with points $t_{i}$ and weights $w_{i}$ depending on the chosen
approximation scheme.

Then
Eq. (\ref{eq2}) becomes a finite linear system of equations, where the kernel has become a matrix.
To symmetrize it we finally introduce
\beq
\tilde{\phi}_i=\frac{v_i}{\sqrt{w_i}},\ \ {\rm or}\ \ \phi(q^2_i)=\frac{1}{\sqrt{q_i^2w_i}}v_i
\label{qv},
\eeq
and the final equation is
\beq
A_i v_i+\sum_{j=0}^N \sqrt{w_iw_j} q_i q_j
 L_{n}(q_i^{2},q_j^{2})v_j=
\epsilon v_i
\label{eq3}
\eeq
with
\beq
A_i=
\sum_{j=0}^N w_j q_j^2\frac{f(q_i)}{f(q_j)}\Big(B_0(q_i^2,q_j^2)-C_0(q_i^2,q_j^2)\Big).
\label{a05}
\eeq

We are going to find the eigenvalues and eigenvectors of the matrix associated to the linear operator of the equation.
The matrix is
\beq
M_{ij}=
A_i\delta_{ij}+\sqrt{w_iw_j} q_i q_j\Big(-B_{n}(q_i^{2},q_j^{2})
+\delta_{n0}\pi\frac{f(0)}{f(q_i)f(q_j)}\Big).
\label{matrix}
\eeq
Its diagonal term is
\beq
M_{ii}=A_i+ w_i q_i^2 \Big(-B_{n}(q_i^{2},q_i^{2})
+\delta_{n0}\pi\frac{f(0)}{f^2(q_i)}\Big).
\label{diag}
\eeq

In order to make explicit the behavior at large momentum, i.e. at  large $t$,
we separate in $A_i$ from the sum over $j$  the term $j=i$ and write
\beq
A_i=
w_iq_i^2\Big(B_0(q_i^2,q_i^2)-C_0(q_i^2,q_i^2)\Big)+ S,
\label{a06}
\eeq
where
\beq
S=\sum_{j=0,j\neq i}w_j q_j^2\frac{f(q_i)}{f(q_j)}\Big(B_0(q_i^2,q_j^2)-C_0(q_i^2,q_j^2)\Big),
\label{s}
\eeq
and combine the separated term with the second term in (\ref{diag}) to obtain
\beq
M_{ii}=w_iq_i^2\Big(B_0(q_i^2,q_i^2)-B_n(q_i^2,q_i^2)-C_0(q_i^2,q_i^2)
+\delta_{n0}\pi\frac{f(0)}{f(q_i)^2}\Big) +S.
\label{diag1}
\eeq
Here the extra large term at large $q_i$ is canceled between $B_0(q_i^2,q_i^2)$ and $B_n(q_i^2,q_i^2)$.
So denoting
\beq
B_{0n}(q_i^2)=B_0(q_i^2,q_i^2)-B_n(q_i^2,q_i^2)
\label{b0n}
\eeq
we find finally
\beq
M_{ii}=w_iq_i^2\Big(B_{0n}(q_i^2)-C_0(q_i^2,q_i^2)
+\delta_{n0}\pi\frac{f(0)}{f(q_i)^2}\Big) +S.
\label{diag2}
\eeq
In this form calculation of eigenvalues of $M$ can be implemented numerically with not much efforts~\footnote{E.g. it can be computed with a FORTRAN program
using only double precision.}.

Note that for the pomeron with $n=0$ the first term in (\ref{diag2}) is zero and we get
\beq
M_{ii}^{n=0}=w_i q_i^2\Big(-C_0(q_i^2,q_i^2)
+\pi\frac{f(0)}{f(q_i)^2}\Big) +S.
\label{diag3}
\eeq
For the odderon (n=1) the last term in (\ref{diag2}) is zero and one has
\beq
M_{ii}^{n=1}=w_iq_i^2\Big(B_{01}(q_i^2)-C_0(q_i^2,q_i^2)\Big) +S.
\label{diag4}
\eeq
In both cases $S$ is given by (\ref{s}).

In this form the matrix $M$ is suitable for the determination of eigenvalues and eigenvectors with standard
numerical techniques. We shall use a grid up to $t_{max}=270$.

\section{Numerical results}
Our numerical results were obtained by the standard FORTRAN programs
for finding all eigenvalues and eigenvectors of a real symmetric matrix.
In the calculations made up to now we have taken
\[ m=0.82 GeV,\ \  \Lambda=0.2 GeV.\]
The lower limit in $t$ was taken in all cases as
\[t_{min}=-90\]
while the upper limit was taken as
\[t_{max}=90,\ 180,\ 270.\]
Moreover the number  of divisions $N$ was taken as
\[N=1000,\ {\rm and},\ 2000.\]

It turned out that the results practically do not change with the rise
of $N$ from 1000 to 2000, so that $N=1000$ proved to be  fairly enough
for a reasonable accuracy. On the other hand, some results do depend on
the value of $t_{max}$. It is worth mentioning that within the used
FORTRAN program we were not able to raise it above 300.

We considered four cases. First we repeated our old calculations for
the pomeron with $n=0$ and the running coupling. Second, we considered
the odderon (n=1) with the running coupling. Finally we studied both the
pomeron and odderon with a fixed coupling regulated by the
"gluon mass" $m$ in the infrared.

\subsection{Pomeron}
For the pomeron we have found that, in agreement with the L.N.~Lipatov picture,
the introduction of the running coupling splits the negative part of the cut
in energy into a sequence of poles converging to the start of the cut at
$E=0$. The number of negative poles rises from 19 at $t_{max}=90$ to
55 at $t_{max}=270$. The negative part of the spectrum at $t_{max}=270$
is shown in Fig. \ref{fig1}. Energies and slopes of the first ten states
at $t_{max}=90$ and $t_{max}=270$ are presented in Table 1. One observes
that their values practically do not depend on $t_{max}$
Note that the first two energies are in agreement with our
old calculations. However their slopes are not, because of a past error in the code which we have corrected.
We can study the energy $\omega_n$ and the slope $\alpha'_n$ of the state $\phi_n$,
as a function of $n$. From a fit on the first $10$ leading states we find for the "energies"
\beq
\omega_n\simeq  -\frac{0.4141}{n+0.1383}\,,
\eeq
which should be compared to the behavior reported in Eq.~\eqref{lipeq3} and which is qualitatively similar also to what found in~\cite{Bartels:2018pin}.
Also the slope can be fitted. It goes roughly as $\alpha'_n=a/(n^2+b)$ for some $a$ and $b$, even if the fit is a bit less accurate compared to what happens for the energies.
\begin{table}
\begin{center}
 \begin{tabular}{||r|r||r|r||}
 \hline
\multicolumn{4}{||c||}{$t_{max}=90$\hspace{3. cm}$t_{max}=270$}\\\hline
     -0.35582E+00&  0.61753E-01&   -0.35584E+00&  0.61731E-01\\
     -0.19280E+00&  0.20731E-01&   -0.19296E+00&  0.20732E-01\\
     -0.13205E+00&  0.10719E-01&   -0.13215E+00&  0.10713E-01\\
     -0.10025E+00&  0.65802E-02&   -0.10066E+00&  0.65780E-02\\
     -0.80711E-01&  0.44542E-02&   -0.80777E-01&  0.44549E-02\\
     -0.67512E-01&  0.32153E-02&   -0.67568E-01&  0.32168E-02\\
     -0.58005E-01&  0.24300E-02&   -0.58053E-01&  0.24317E-02\\
     -0.50834E-01&  0.19017E-02&   -0.50878E-01&  0.19025E-02\\
     -0.45214E-01&  0.15478E-02&   -0.45275E-01&  0.15290E-02\\
     -0.40496E-01&  0.14027E-02&   -0.40780E-01&  0.12555E-02\\\hline
\end{tabular}
\end{center}
\caption{Energies and slopes of the first 10 states of the pomeron
with the running coupling}
\end{table}

\begin{figure}
\centerline{\includegraphics[width=10cm]{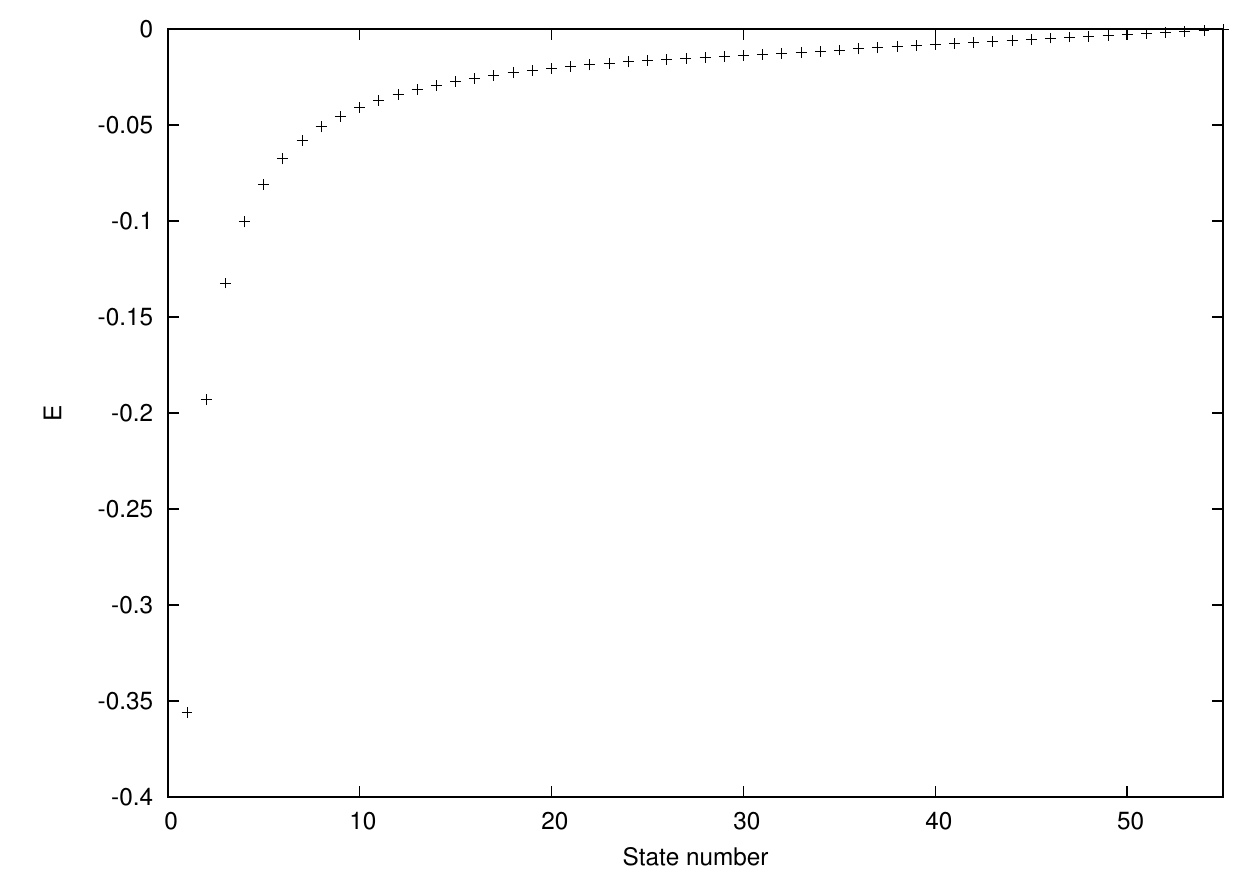}}
\caption{Energies of the first 55 states of the
pomeron with the running coupling. $t_{max}=270$}
\label{fig1}
\end{figure}

\begin{figure}
\centerline{\includegraphics[width=10cm]{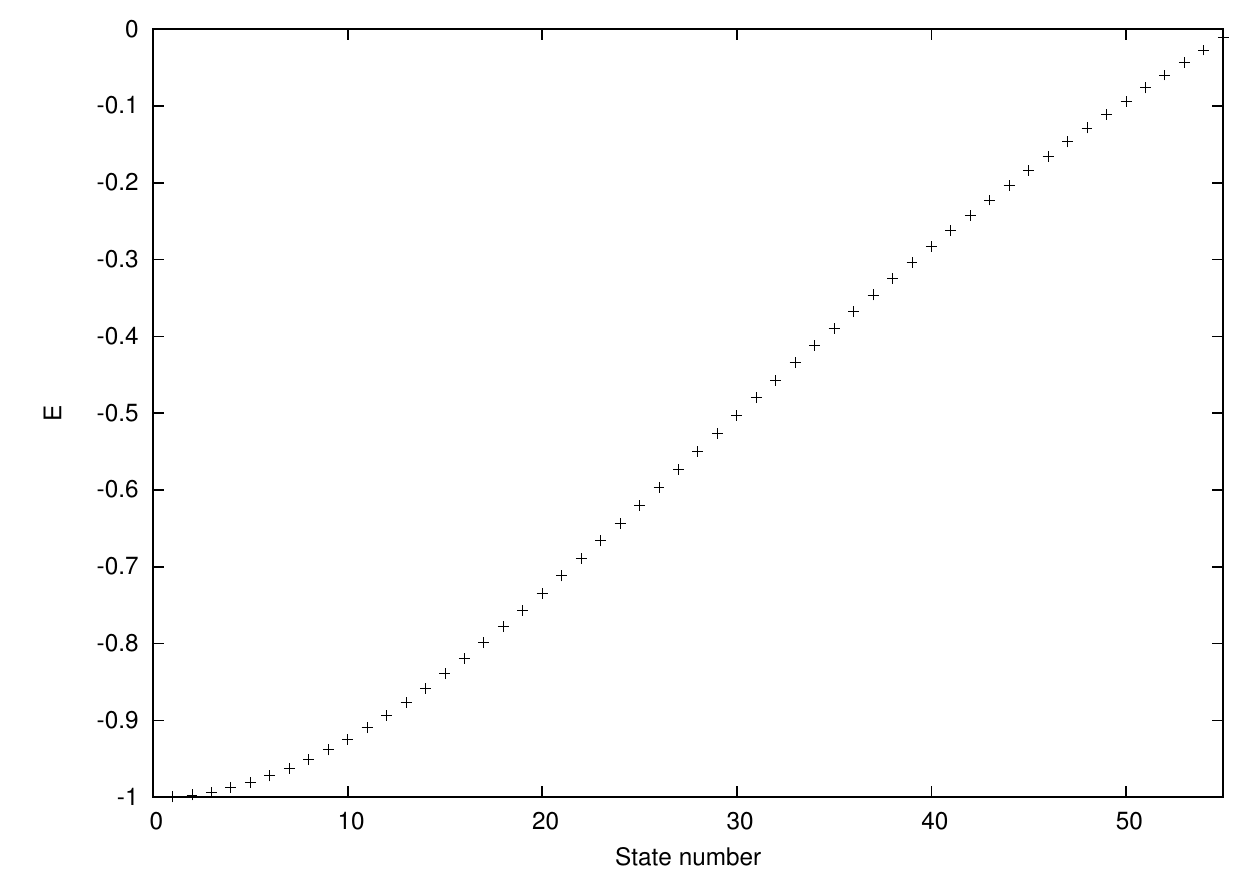}}
\caption{Energies of the first 55 states of the
pomeron with the fixed coupling. $t_{max}=270$}
\label{fig2}
\end{figure}

To compare we calculated energies (in units $|E_{BFKL}|$) and slopes
for the pomeron with a fixed coupling. Here we observe a cut in energy
starting at exactly the BFKL endpoint and going upwards, as predicted
by previous calculation. Remarkably the number of negative states is the
same as for the running coupling: 19 with $t_{max}=90$ and 55 with
$t_{max}=270$. However the location of energies is quite different
indicating a cut, as illustrated in Fig. \ref{fig2}.
Also values of the intercept were found to correspond to a non-normalizable state.
They are not zero but strongly diminish with the growth of $t_{max}$,
which can be seen from Table 2. Such behavior $\propto 1/t_{max}$ is
of course expected on general grounds.

\begin{table}
\begin{center}
 \begin{tabular}{||r|r||r|r||}
 \hline
\multicolumn{4}{||c||}{$t_{max}=90$\hspace{3. cm}$t_{max}=270$}\\\hline
     -0.99335E+00&  0.25232E-03&   -0.99953E+00&  0.10250E-04\\
     -0.97347E+00&  0.98302E-03&   -0.99714E+00&  0.40871E-04\\
     -0.94143E+00&  0.21184E-02&   -0.99318E+00&  0.91482E-04\\
     -0.89868E+00&  0.35509E-02&   -0.98767E+00&  0.16146E-03\\
     -0.84704E+00&  0.51579E-02&   -0.98064E+00&  0.24995E-03\\
     -0.78851E+00&  0.68197E-02&   -0.97213E+00&  0.35588E-03\\
     -0.72505E+00&  0.84346E-02&   -0.96219E+00&  0.47801E-03\\
     -0.65850E+00&  0.99274E-02&   -0.95088E+00&  0.61492E-03\\
     -0.59048E+00&  0.11252E-01&   -0.93826E+00&  0.76509E-03\\
     -0.52233E+00&  0.12388E-01&   -0.92439E+00&  0.92686E-03\\\hline
\end{tabular}
\end{center}
\caption{Energies and slopes of the first 10 states of the pomeron
with the fixed coupling}
\end{table}

\begin{figure}[ht]
\centerline{\includegraphics[width=10cm]{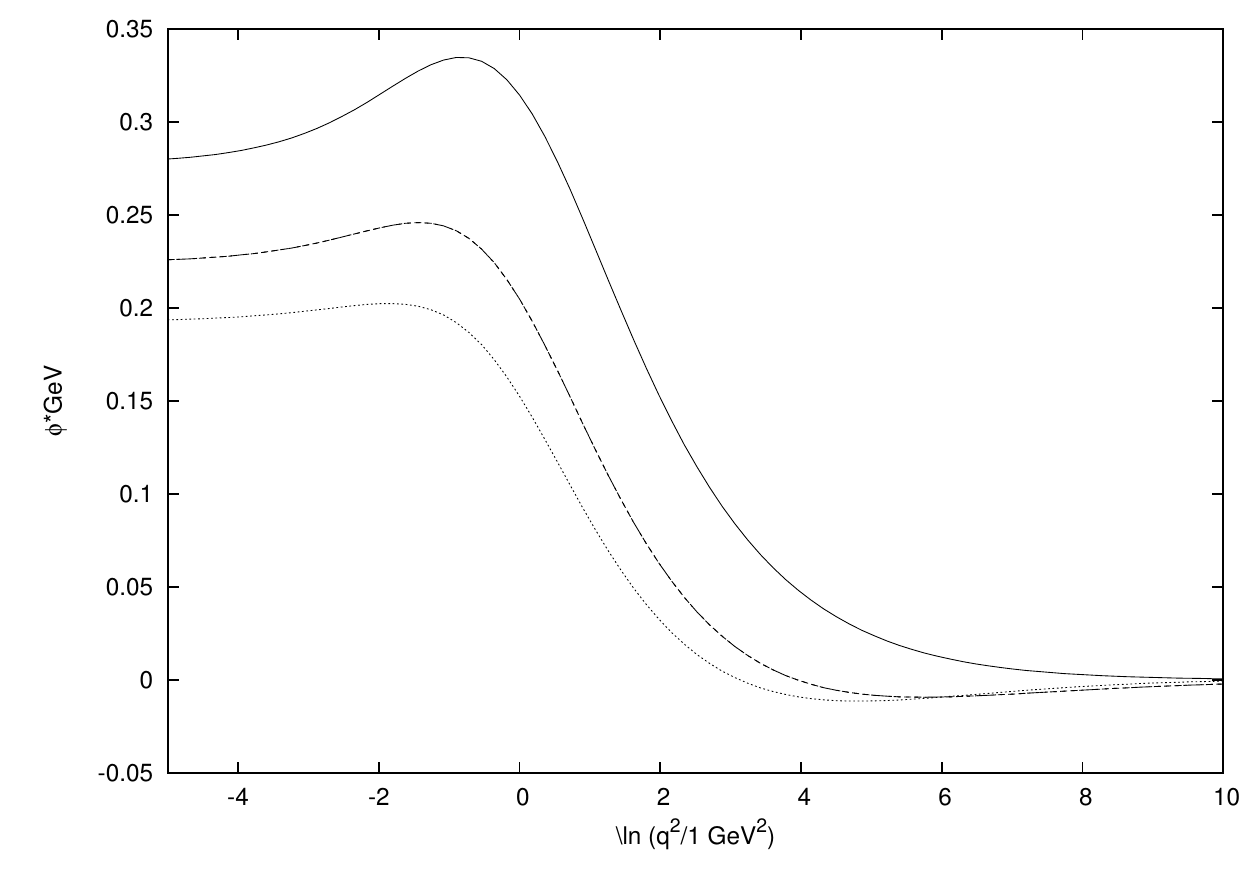}}
\caption{Wave functions $\phi_n(q^2)$ as functions of $q^2$ for the first three
states  with the lowest energy. From top to bottom curves correspond to the number of state
$n=1,2,3$}
\label{fig5}
\end{figure}

We also studied the behavior of the pomeron wave functions
$\phi(q^2)$ as functions of $q^2$ for the three states with the lowest energy.
It is shown in Fig. \ref{fig5}. Remarkably all three wave functions are very similar
in their $q^2$ dependence in the low momentum region, but at larger momenta different states have a different number of nodes (zeros).

\begin{figure}[ht]
\centerline{\includegraphics[width=4cm]{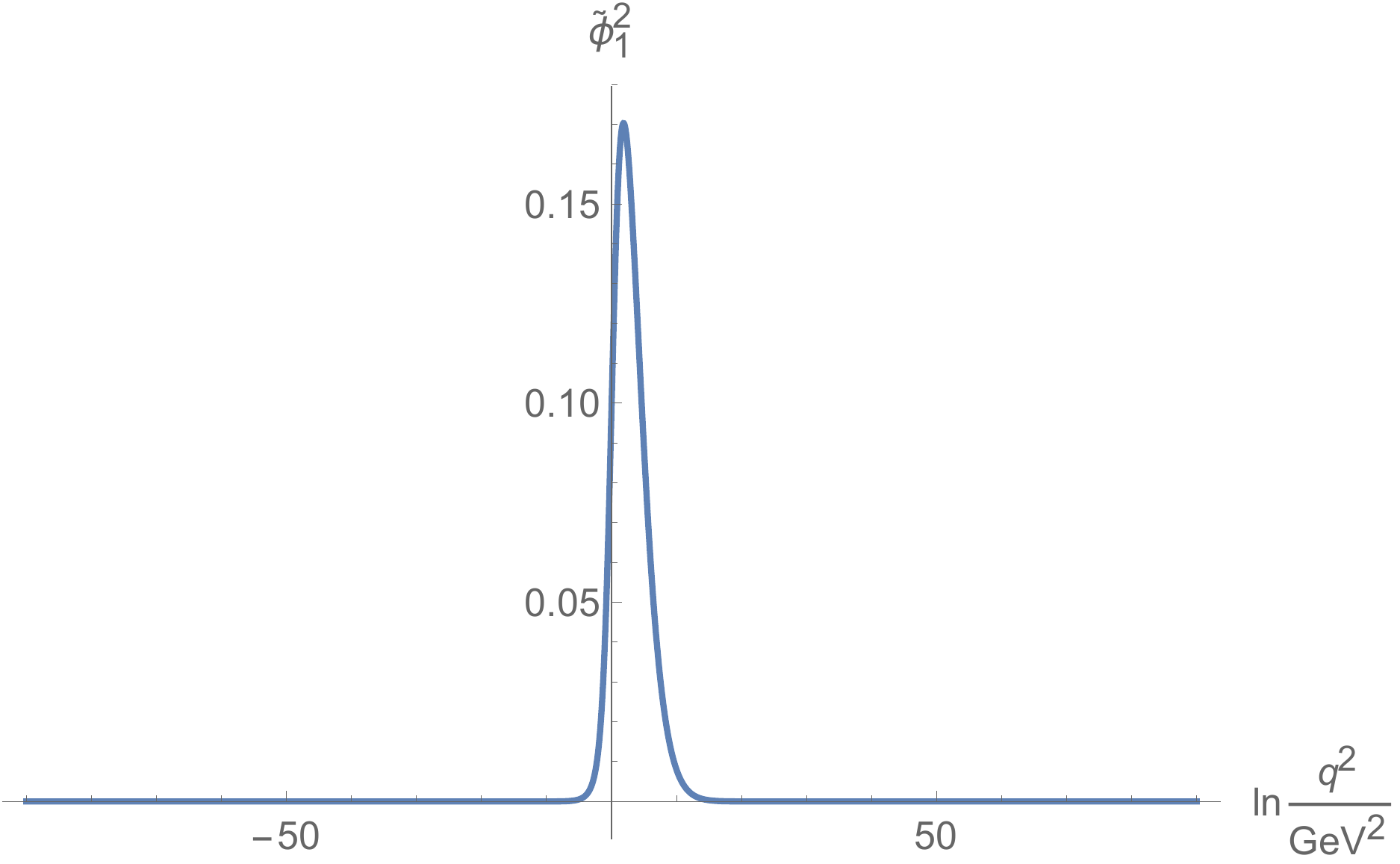}\hspace{0.3cm}
\includegraphics[width=4cm]{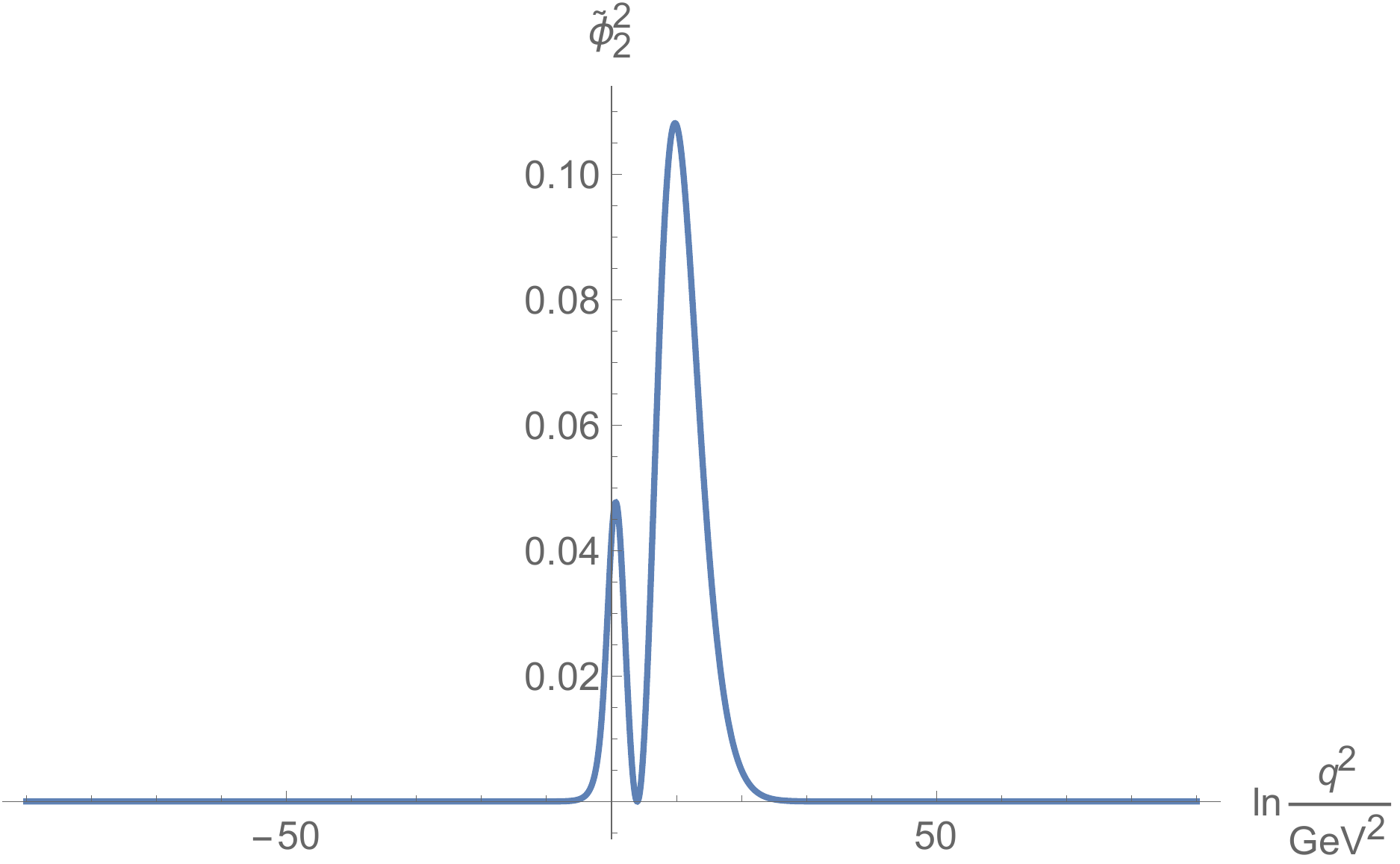}\hspace{0.3cm}
\includegraphics[width=4cm]{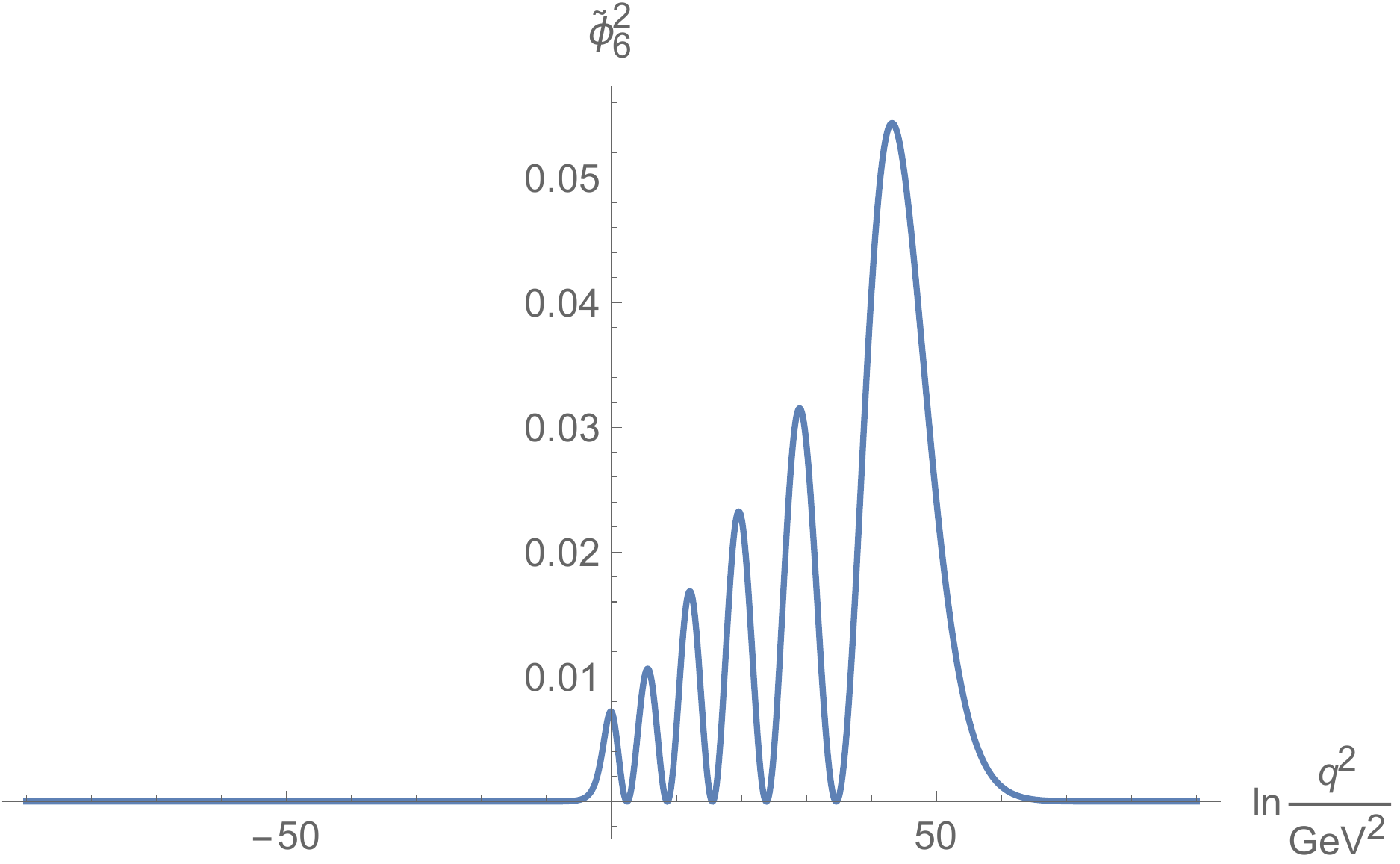}}
\caption{Square of the pomeron eigenfunctions $\tilde{\phi}_i(q^2)$ for the first, second and sixth leading energy states.
The region of highest momentum support is pushed rapidly towards the UV with the increasing order of the eigenstate.}
\label{figsupp}
\end{figure}

Let us analyze how the support in momentum space of the discrete states differ. This is best understood analyzing the rescaled eigenfunctions $\tilde{\phi}_j(q^2)$.
The square of the first, second and sixth states with lowest energies are shown in Fig~\ref{figsupp}.
This behaviour is similar to the one found in ~\cite{kowalski2} where excited states show a clear shift towards higher values of $q^2$.
One can see the dependence on the state number $\tilde{\phi}_j(q^2)$ of the position $q_{UV}(j)$ of the center of the most UV region of support. The dependence is very well approximated with a simple linear fit. For example taking the first ten states one gets
\beq
\ln q_{UV}^2(j)/{\rm GeV^2} \simeq  -7.278 + 8.456  j.
\eeq

Finally let us comment on the dependence on the mass parameter m which also appears inside the logarithmic dependence of the running coupling
(keeping the same value of $\Lambda_{QCD}$), and whose value fixes the order of the scale where $\alpha_s$ feezes. Lowering $m$ from $0.82$ GeV to $0.52$ Gev, for the leading state the  energy ($=1-$ intercept) increases in absolute value of about $20\%$ while the slope changes much more, becoming almost $4$ times larger, which is close to the well known value $0.25GeV^{-2}$.

\subsection{Odderon}

\begin{figure}
\centerline{\includegraphics[width=10cm]{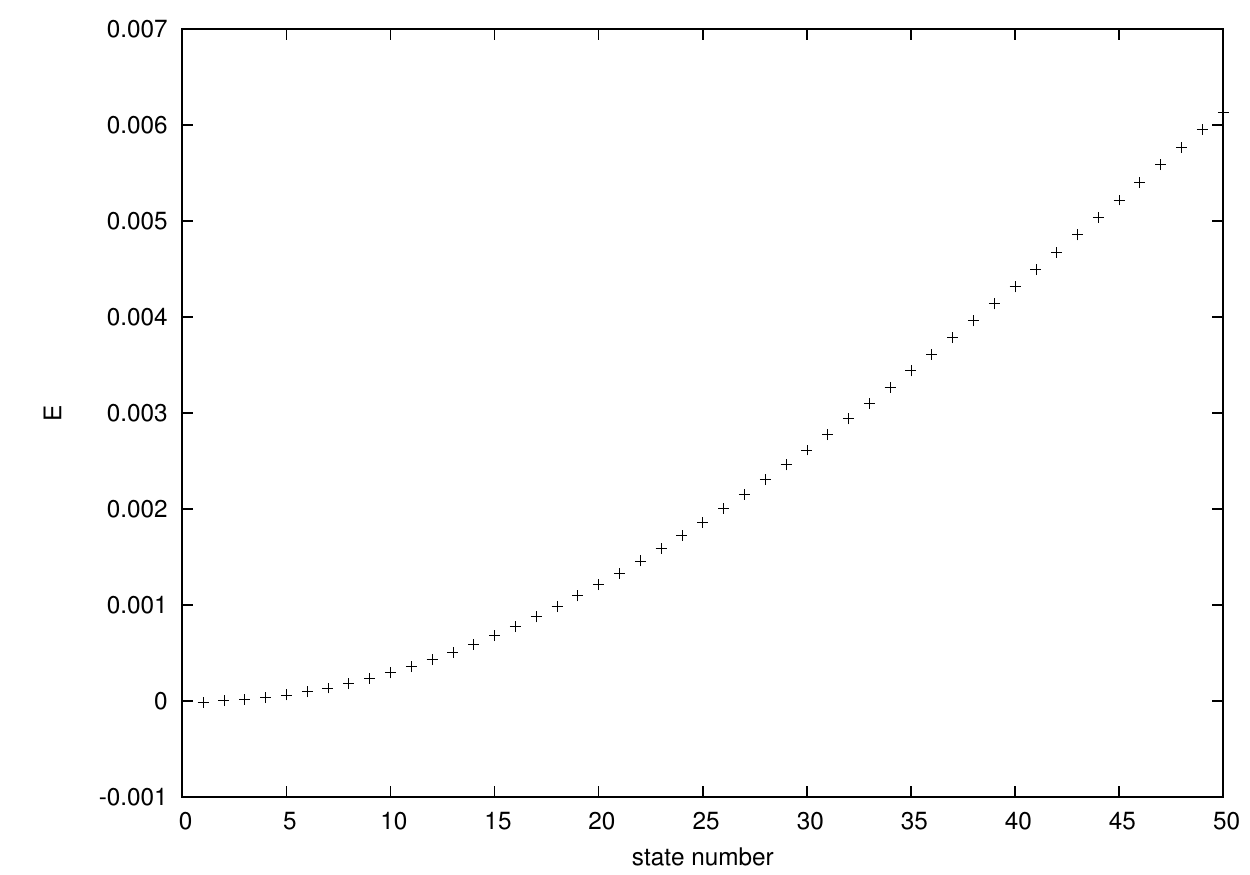}}
\caption{Energies of the first 50 states of the
odderon with the running coupling. $t_{max}=270$}
\label{fig3}
\end{figure}

\begin{figure}
\centerline{\includegraphics[width=10cm]{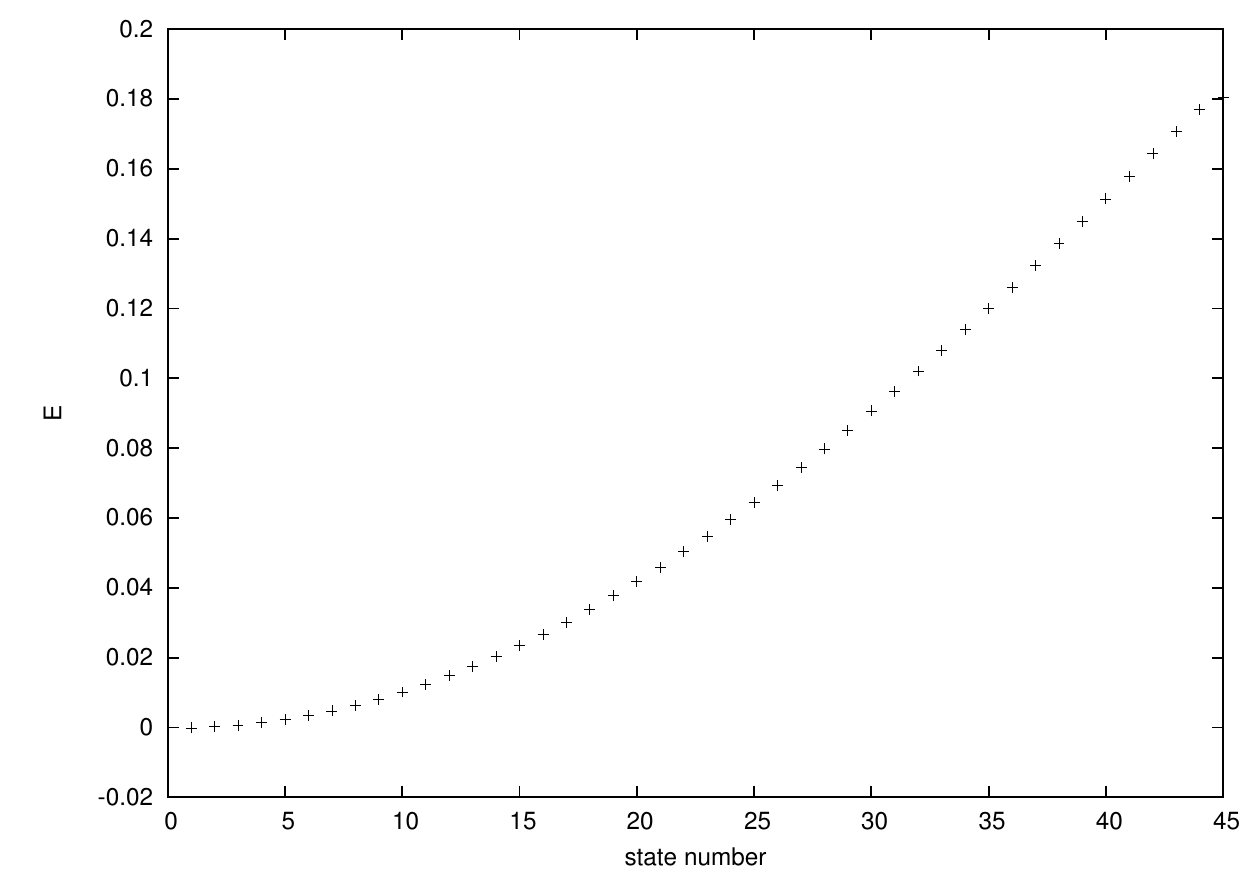}}
\caption{Energies of the first 55 states of the
odderon with the fixed coupling. $t_{max}=270$}
\label{fig4}
\end{figure}

Passing to the odderon let us first discuss the intercept.
We have found that both with the running and
fixed coupling constants energies lie on the cut starting at $E=0$.
The density of states rises with $t_{max}$. In particular with the
running coupling the number of
states with $E<0.02$ rises from 18 at $t_{max}=90$ to 135 at
$t_{max}=270$. With the fixed coupling constant the density of states
is considerably smaller and rises not so fast. At $t_{max}=90$ we found
only 5 states with $E<0.02$ and at  $t_{max}=270$ their number rises to
13. Location of energies smaller than 0.02 for low-lying states for
the running and fixed coupling constants is illustrated in Figs.
\ref{fig3} and \ref{fig4} respectively.

\begin{figure}[ht]
\centerline{\includegraphics[width=10cm]{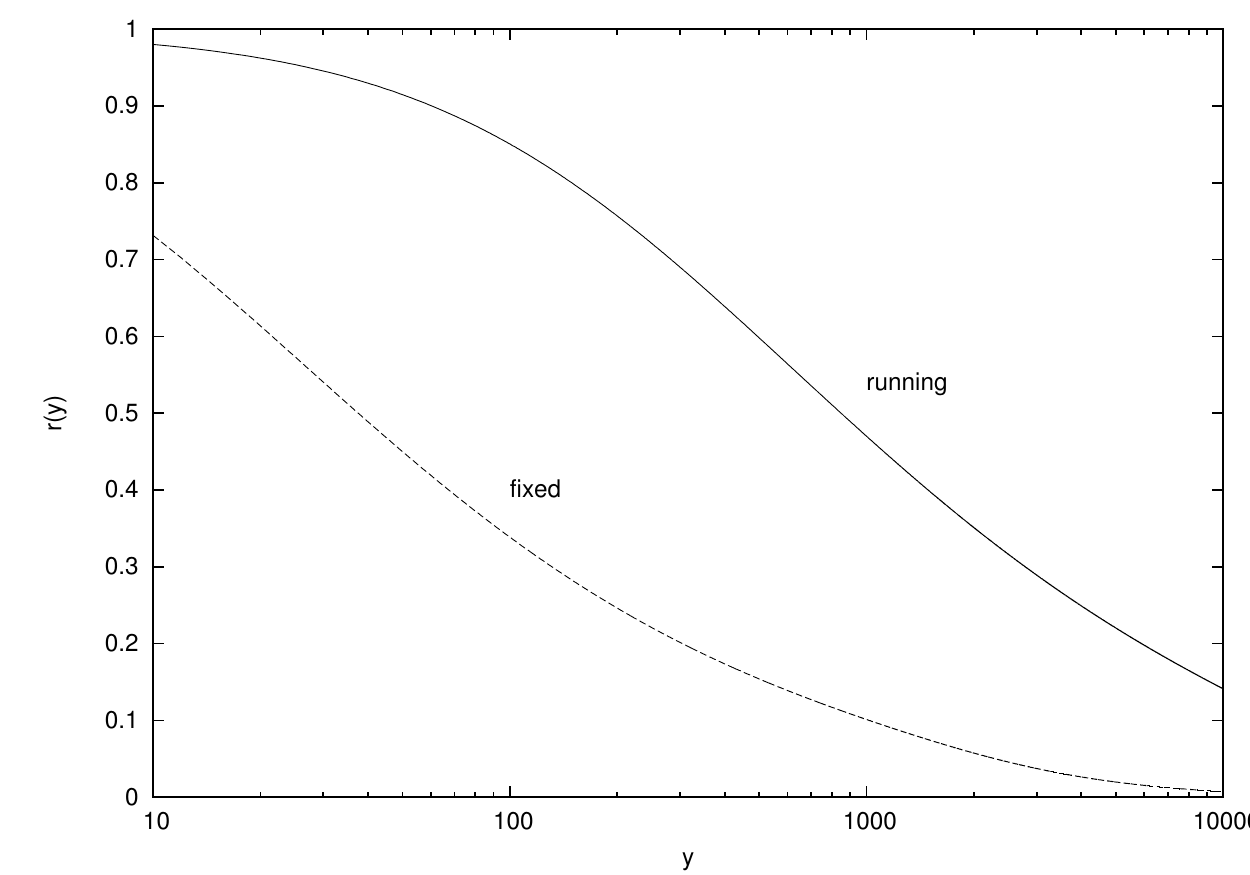}}
\caption{The behavior of the amplitude with the odderon exchange $A(y)$ as a function of rapidity $y$.
The upper and lower curves show the ratios $A(y)/A(0)$ with and without running of the coupling}
\label{fig-odd-amp}
\end{figure}

For the odderon we studied the character of the singularity of its spectrum at the
branch-point $E=0$. It is revealed in the behavior of the amplitude governed by the
odderon exchange at high $s$, Namely if the spectral density $\tau(E)$ has a singularity
$E^\beta$ at $E=0$ the amplitude will diminish as $y^{-(1+\beta)}$ at large $y=\ln s$.
In our calculations the continuous spectrum is approximated by a set of poles at $E_i$,
$i=1,2,3,...$ with very small distances from one another. We introduced the coupling $a_i$ of
the $i$-th odderon state to the projectile or target taking for each of them a certain distribution
of colour $\rho(q^2)=c\exp(-r_0q^2)$ with $r_0$ of the order of the proton radius.  Then
\beq
a_i=\sum_j\sqrt{w_jq_j}\rho(q_j^2)v_j^{(i)}\nonumber
\eeq
where $v^{(i)}$ is the eigenvector for the $i$-th odderon (actually its wave function on the grid).
The discrete spectral density is accordingly taken as $\tau_i=a_i^2$.
With this we studied the $y$-dependence of the  amplitude
coming from the exchange of all odderon states
\beq
{\cal A}(y)=\sum_ia_i^2e^{-E_iy}\,.
\eeq
The ratios $r(y)={\cal A}(y)/{\cal A}(0)$ are plotted in Fig. \ref{fig-odd-amp} for the odderon with running and fixed
coupling constants. It is clearly seen that  running of the coupling makes the amplitude fall somewhat weaker at high $y$.
This shows how important is important the incusion of running coupling effects for any quantitative estimate of an odderon exchange cross section.
Fits to the curves give the behavior $\sim y^{-0.38}$ and $\sim y^{-0.86}$ for the running and fixed coupling respectively.
This implies that the singularity at $E=0$ of the spectral density is  stronger
with the running coupling $E^{-0.62}$ than with the fixed one  $E^{-0.14}$.

These results are qualitatively similar to the ones recently found using another prescription to introduce the running coupling~\cite{Bartels:2019qho}.

\section{Conclusions}
We have focused on some aspects of QCD strong interactions involving the so called odderon exchange,
which may be relevant in the scattering process in the Regge limit and also in presence of dense nuclear media,
and are still posing challenging problems both theoretically and experimentally.
It is still to be understood and investigated more at experimental level the difference at high energies in the scattering of $pp$ and $p\bar{p}$,
even if new data started to arrive from the TOTEM experiment.
Non perturbative physics and the uncertainties in non linear effects on parton distributions are main theoretical obstacles.

We have discussed, in the context of perturbative QCD and BFKL physics, a path to improve the description of a family of known and leading odderon solutions (BLV) as well as more general BKP states in the LL approximation and large $N_c$ limit to account for running coupling effects. This is obtained requiring a match with standard one loop running coupling properties which leaves the freedom to model the IR non perturbative region and at the same time respects s-channel unitarity for the reggeization of the gluon at the base of the BFKL approach. This so called bootstrap approach was shown at perturbative level to reproduce running coupling properties of the NLL BFKL. For the pomeron channel we explicitly show that this procedures leads to a series of discrete leading poles with intercept greater than one, which accumulates at one.

For the odderon the inclusion of the running coupling makes more dense the states close to intercept one, where the cut continues to start.
Therefore the rapidity dependence in the scattering amplitudes is affected making cross section larger, so that running coupling effects should not be neglected.

Pomeron and odderon fields can be elements of an effective description to study rapidity dependence and saturation effects in dense nuclear media or in extreme high energy scattering processes. Alternative descriptions to the reggeized gluon approach are the color dipole picture (large $N_c$) and the CGC approach.
We have shown how to reproduces in the BFKL description the non linear evolution of the BFKL pomeron induced by the odderon splitting
which was known from previous CGC studies.
To derive in the BFKL framework also a non linear evolution for the odderon~\cite{hatta} the $V_{3\to5}$ reggeized gluon vertex should be computed.
One of the main theoretical problems is that these effective theories do not take into account pomeron/odderon loops since only tree level effective equations are available.
The loops were formally considered in the old days before the advent of QCD in the effective reggeon field theory where local fields for pomeron and odderon were considered.
This description may be valid only for very large transverse distances.
We have recalled how one can compute the critical properties of such a theory and extract the universal behavior (critical exponents) even if this regime
if far from be probed by any experimental setup.
It is interesting that there is a universal link with statistical mechanical systems, in particular with the out of equilibrium dynamics in generalized directed percolating systems.

Finally we note that in the LL generalized BFKL framework all the known gauge invariant effective vertices are build using the 
bootstrap condition, and involve always reggeized gluon trajectories and the real production part of the two gluon BFKL kernel. 
Therefore one can envisage a way to introduce also there the running coupling by the same procedure to take into account this part of the 
effects which would otherwise appear only at the level of the NLL approximation.

To conclude we want to stress the most important and seminal contributions made by Lev Lipatov in this field and discusses here. 
Apart from being one of the father-founders of the whole reggeized gluons approach, he specifically actively dealt with odderon and BKP
states during his career with outstanding results.
He studied the conformal invariance of the odderon wave function and formulated equations which later were used to find the JW-odderon.
He discovered that in the planar (multicolor) limit the BKP equations become fully integrable and showed their relation to the non-compact Heisenberg spin chain,
 which formed the basis of the subsequent study of their properties. This was the first demonstration of integrability in gauge theories.
Lev Lipatov was the first to discuss the influence of the running of the coupling and to conclude that it splits the negative cut in the energies
of the pomeron into a sequence of poles, the results fully confirmed later irrespective of the way to introduce this running.
Lev Lipatov was one of the authors to show the equivalence of the reggeized gluons  and dipole approaches for the fan diagram BK evolution equation
and  in relation to this formulated the correct transition to the coordinate representation as the transition to  "the  M\"obius representation".
He was also one of the authors to investigate the next-to-leading order for the odderon structure, the result which will certainly play its role in future studies.

\section{Appendix. The slope}

The slope can be obtained starting from Eq.~(\ref{basic}), expanding up to second order in $q$ the relation
\beq
\langle (h_q-h_0) \rangle_0 = \epsilon(q)-\epsilon(0)\,,
\eeq
where the average is done on the eigenstates in the forward direction ($q=0$) and
according to the previous notation as in Eq.~(\ref{lq}) we consider separate contributions and write
\beq
h_q-h_0=W^{kin}(q)+W^{ql}(q)+W^{sep}(q)\,.
\eeq
First we consider the function $f(p)$ of Eq.~(\ref{eta1}) which is a kind of buiding block appearing in different instances.  According to the decomposition
of Eq.~(\ref{qll}) in the first two orders of $q$ we get
\beq
f(q_1)=f(l)\Big[1+a_1\Big((ql)+\frac{1}{4}q^2\Big)+\frac{1}{2}a_2(ql)^2\Big],
\label{fq1}
\eeq
where we denote with $ql=q\cdot l$ the scalar product of the two dimensional vectors, and
\beq
a_1=\frac{1+\ln[(l^2+m^2)/\Lambda^2]}{f(l)},\ \ a_2=\frac{1}{(l^2+m^2)f(l)}.
\label{a12}
\eeq
From this result one directly finds also $f(q_2)$, $f(q'_1)$ and $f(q'_2)$ simply changing
$l\to -l$, $l\to l'$ and $l\to -l'$ respectively. Coefficients $a_1(l)$ and $a_2(l)$ do not change
under $l\to -l$. As in ~\cite{bvv} we denote them simply as $a_1$ and $a_2$. Coefficients $a_1(l')$
and $a_2(l')$ will be denoted $a'_1$ and $a'_2$
Also
\beq
f(q)=f(0)(1+a_3 q^2),\ \ a_3=\frac{1+\ln m^2/\Lambda^2}{f(0)}\,.
\label{fqq}
\eeq
Note also that $q_1-q'_1=q_2'-q_2=l-l'$ do not depend on $q$.

\subsection{$W^{kin}$}
The kinetic term (\ref{a0q}) contains two terms corresponding to trajectories $\omega_1$ and $\omega_2$.
For the first part of $W^{kin}$ coming from $\omega_1$ we find
\[
W^{kin}_1=\frac{1}{2}\int d^2l'\frac{f(l)}{f(l')f(l-l')}
\Big[
a_1\Big((ql)+\frac{1}{4}q^2\Big)+\frac{1}{2}a_2(ql)^2\]\beq
-a'_1\Big((ql')+\frac{1}{4}q^2\Big)+\Big(a'_1{}^2-\frac{1}{2}a'_2\Big)(ql')^2
-a_1a'_1(ql)(ql')\Big].
\label{wkin1}
\eeq
The second part of $W^{kin}$ is found to be
\[
W^{kin}_2=\frac{1}{2}\int d^2l'\frac{f(l)}{f(l')f(l-l')}
\Big[
a_1\Big(-(ql)+\frac{1}{4}q^2\Big)+\frac{1}{2}a_2(ql)^2\]
\beq
-a'_1\Big(-(ql')+\frac{1}{4}q^2\Big)+\Big(a'_1{}^2-\frac{1}{2}a'_2\Big)(ql')^2
-a_1a'_1(ql)(ql')\Big].
\label{wkin2}
\eeq
In the total $W^{kin}$ terms linear in $q$ cancel and quadratic ones are doubled. Thus at second order in $q$ we find
\beq
W^{kin}=\int d^2l'\frac{f(l)}{f(l')f(l-l')}
\Big[
\frac{1}{4}q^2(a_1-a'_1)+\frac{1}{2}a_2(ql)^2
+\Big(a'_1{}^2-\frac{1}{2}a'_2\Big)(ql')^2
-a_1a'_1(ql)(ql')\Big].
\label{wkin3}
\eeq

\subsection{$W^{ql}$}
At finite $q$ this part is given by (\ref{lql}). One finds at second order in $q$
\beq
\sqrt{\frac{f(q_1)}{f(q_2)}}=1+a_1(ql)+\frac{1}{2}a_1^2(ql)^2,\ \
\sqrt{\frac{\fd}{\fc}}=1-a'_1(ql')+\frac{1}{2}{a'_1}^2(ql')^2.\nonumber
\eeq
and taking the product
 \beq
 \sqrt{\frac{f(q_1)\fd}{f(q_2)\fc}}=
 1+a_1(ql)-a'_1(ql')+\frac{1}{2}\Big(a_1(ql)-a'_1(ql')\Big)^2. \nonumber
 \eeq

 In the second term of $L$ in Eq.~(\ref{lql}) we have to change $l\to -l$ and $l'\to -l'$.
 So the total $q$-dependent factor becomes
 \[
 2+\Big(a_1(ql)-a'_1(ql')\Big)^2
 \]
 from which one finally reads at second order in $q$
 \beq
 W^{ql}(l,l')=-\frac{\Big(a_1(ql)-a'_1(ql')\Big)^2}{f(l-l')}.
 \label{wkin}
 \eeq

 \subsection{$W^{sep}$}
 The separable part of the interaction is given by (\ref{lsep}).
 Here one obtains in the straightforward manner
\[
\frac{f(q)}{\sqrt {f(q1)f(q_2)f(q'1)f(q'_2)}}\]\[
=\frac{f(0)}{f(l)f(l')}
\Big[1+q^2\Big(a_3-\frac{1}{4}(a_1+a'_1)\Big)-\frac{1}{2}(a_2-a_1^2)(ql)^2-\frac{1}{2}(a'_2-{a_1}'^2)(ql')^2\Big].\]
From this one gets at second order in $q$
\beq
W^{sep}(l,l')=\frac{f(0)}{f(l)f(l')}
\Big[q^2\Big(a_3-\frac{1}{4}(a_1+a'_1)\Big)-\frac{1}{2}(a_2-a_1^2)(ql)^2-\frac{1}{2}(a'_2-{a_1}'^2)(ql')^2\Big].
\label{wsep}
\eeq

\subsection{Averaging in the state with $n=0$}

First task is to fix the normalization.
Let
\beq
{\cal N}=\int d^2q\, \phi^2(q^2)=\pi\int dq^2\phi^2(q^2)=\pi\int dt q^2\phi^2(q^2).
\eeq
On the grid we get
\beq
{\cal N}=\pi\sum_{i=0}^Nw_iq_i^2\phi^2(q_i^2)=\pi\sum_{i=0}^Nv_i^2.
\eeq
where we used (\ref{qv}). The normalization can be fixed requiring
\beq
\sum_{i=0}^Nv_i^2=1
\eeq
in accordance with the standard programs for the search of eigenvalues and eigenvectors.
This implies
$
{\cal N}=\pi
$
and means that after the integration of  our $W$ with the functions $\phi(q^2)$
one has to divide the result by $\pi$.

{\bf 1. $\langle W^{kin}\rangle$}.

In this way we find
\[
\langle W^{kin} \rangle=\frac{1}{4}q^2\int dl^2\phi^2(l)
\int d{l'}^2\frac{f(l)}{f(l')}\]
\beq
\times
\Big[
\Big(\frac{1}{2}(a_1-a'_1)+\frac{1}{2}a_2l^2
+(a'_1{}^2-\frac{1}{2}a'_2) {l'}^2\Big)B_0(l^2,{l'}^2)
-a_1a'_1 ll' B_1(l^2,{l'}^2)\Big].
\label{avwkin1}
\eeq
Passing first to the integration over $t$ and $t'$ and to summation on the grid we get
\[
\langle W^{kin} \rangle=\frac{1}{4}q^2\sum_{i,j}\phi^2(l_i)l_i^2l_j^2 w_iw_j
\frac{f(l_i)}{f(l_j)}\]\beq\times
\Big[
\Big(\frac{1}{2}(a_{1i}-a_{1j})+\frac{1}{2}a_{2i}l_i^2
+(a_{1j}^2-\frac{1}{2}a_{2j}) l_j^2\Big)B_0(l_i^2,l_j^2)
-a_{1i}a_{1j} l_il_j B_1(l_i^2,l_j^2)\Big].
\label{avwkin3}
\eeq
The final expression  in terms of $v$ is obtained after using (\ref{qv})
\[
\langle W^{kin}\rangle=\frac{1}{4}q^2\sum_{i,j}v_i^2l_j^2 w_j
\frac{f(l_i)}{f(l_j)}\]\beq\times
\Big[
\Big(\frac{1}{2}(a_{1i}-a_{1j})+\frac{1}{2}a_{2i}l_i^2
+(a_{1j}^2-\frac{1}{2}a_{2j}) l_j^2\Big)B_0(l_i^2,l_j^2)
-a_{1i}a_{1j} l_il_j B_1(l_i^2,l_j^2)\Big].
\label{avwkin4}
\eeq

{\bf 2. $\langle W^{ql}\rangle$}

Doing the angular integrations and dividing by $\pi$ we obtain
\[
\langle W^{ql}\rangle=
-\frac{1}{4}q^2\int dl^2 d{l'}^2\phi(l)\phi(l')
\Big[\Big(a_1^2l^2+a'_1{}^2{l'}^2\Big)B_0(l^2,{l'}^2)-2a_1a'_1ll'B_1(l^2,{l'}^2)\Big].
\]
In going to variables $t$ and $t'$ and then to the summation on the grid, using  (\ref{qv})
\[
\langle W^{ql}\rangle=
-\frac{1}{4}q^2\sum_{i,j} \sqrt{w_iw_j}l_il_jv_iv_j
\Big[\Big(a_{1i}^2l_i^2+a_{1j}^2l_j^2\Big)B_0(l_i^2,l_j^2)-2a_{1i}a_{1j}l_il_jB_1(l_i^2,l_j^2)\Big].
\]

{\bf 3. $\langle W^{sep}\rangle$}

Starting from the expression~(\ref{wsep}) we have
\[
\langle W^{sep}\rangle=
\frac{1}{4\pi}\int dl^2d{l'}^2\phi(l)\phi(l')d\chi d\chi'
\frac{f(0)}{f(l)f(l')}\]\[\times
\Big[q^2\Big(a_3-\frac{1}{4}(a_1+a'_1)\Big)-\frac{1}{2}(a_2-a_1^2)(ql)^2-\frac{1}{2}(a'_2-{a'_1}^2)(ql')^2\Big].\]
Here all the angular dependence is trivial. After angular integration terms no depending on the angles give $4\pi^2$
and those containing $\cos^2\chi$ or $\cos^2\chi'$ give $2\pi^2$. So we find
\[
\langle W^{sep}\rangle=
\pi q^2  \!\!\int \! dl^2d{l'}^2
\frac{\psi(l)\psi(l') f(0)}{f(l)f(l')}
\Big[a_3-\frac{1}{4}(a_1+a'_1)-\frac{l^2}{4}(a_2-a_1^2)-\frac{{l'}^2}{4}(a'_2-{a'_1}^2)\Big].\]
Passing first to integration over $t$ and $t'$ and then on the grid we finally find
\[
\langle W^{sep}\rangle=
\pi q^2\sum_{i,j}\sqrt{w_iw_j}
\frac{ v_i v_j l_i l_j f(0)}{f(l_i)f(l_j)}
\Big[a_3-\frac{1}{4}(a_{1i}+a_{1j})-\frac{l_i^2}{4}(a_{2i}-a_{1i}^2)-\frac{l_j^2}{4}(a_{2j}-a_{1j}^2)\Big].\]

The final rescaled slope is given by
\[\tilde{\alpha}'=\frac{1}{q^2}\langle W^{kin}+W^{ql}+W^{sep}\rangle.\]

\subsection{Fixed coupling}
For comparison it is instructive to consider the simple case of the fixed coupling with regularization in the
infrared:
\beq
f(q)=q^2+m^2.
\label{ffix}
\eeq
The energy $E$ is then related to $\epsilon$ as
\beq
E=\frac{3\alpha_s}{2\pi^2}\epsilon,
\eeq
where $\alpha_s$ is the fixed coupling constant. To eliminate dependence on it is convenient
to measure its relation of the energy of the BFKL ground state  $E_{BFKL}=12\alpha_s\ln2/\pi$.
Then
\beq
E=\epsilon\frac{\epsilon}{8\pi\ln 2}E_{BFKL}.
\eeq

With (\ref{ffix}) functions $B_n$ and $C_0$ become known analytically.
In particular,
\beq
B_0(q_1^2,q_2^2)=\frac{2\pi}{r},\ \ r^2=(q_1^2-q_2^2)^2+2m^2(q_1^2+q_2^2)+m^4,
\eeq
\beq
B_1(q_1^2,q_2^2)=\frac{2\pi b}{r(a+r)},\ \ a=q_1^2+q_2^2+m^2,\ \ b=2q_1q_2,
\eeq
\beq
B_{01}(q_1^2,q_1^2)=\frac{2\pi}{b}\Big(1-\frac{r}{a+b}\Big)
\eeq
and
\beq
C_0(q2,q_1^2)=\frac{2\pi}{r_1},\ \ r_1^2=(q^2+2q_1^2+2m^2)^2-4q^2q_1^2.
\eeq

In the calculation of slopes we find
\beq
a_1(q)=\frac{1}{f(q)},\ \ a_2=0,\ \ a_3=\frac{1}{f(0)}\,.
\eeq









\begin{thebibliography}{99}


\bibitem{ewerz} C.Ewerz, (2003) arXiv: hep-ph/0306177.
\bibitem{lukanic}  L.Lukaszuk and B.Nicolescu, \emph{Lett. Nuovo Cim.} {\bf 8}  405 (1973).
\bibitem{bfkl1} E.A.Kuraev, L.N.Lipatov, V.S.Fadin, \emph{Sov. Phys. JETP} {\bf 45}  199 (1977).
\bibitem{bfkl2} I.I.Balitski, L.N.  Lipatov, \emph{Sov. J. Nucl. Phys.} {\bf 28} 822  (1978).
\bibitem{lipfadin}V.S.Fadin, L.N.Lipatov, \emph{Phys. Lett. B} {\bf 429} 127  (1998).
\bibitem{ciafa}M.Ciafaloni, G.Camici, \emph{Phys. Lett. B} {\bf 430}  349 (1998), [arXiv: hep-ph/9803389].
\bibitem{barwue} J.Bartels, M.Wuesthoff, \emph{Z. Physik C} {\bf 66}  157 (1995).
\bibitem{mueller} A.Mueller, B.Patel, \emph{Nucl. Phys. B} {\bf 425}  471 (1994).
\bibitem{bal} I.Balitski, \emph{Nucl. Phys. B} {\bf  463}  99 (1996).
\bibitem{kov} Yu.V. Kovchegov, \emph{Phys. Rev. D}  {\bf 60} 034008  (1999).
\bibitem{iancu}E.Ferreiro, E.Iancu, A.Leonidov, L. McLerran; \emph{Nucl. Phys.A} {\bf 703} 489  (2002).
\bibitem{bra}M. Braun \emph{Eur. Phys. J. C} {\bf 16} (2000) [hep-ph/0001268].
\bibitem{blv1}
  J.~Bartels, L.~N.~Lipatov and G.~P.~Vacca,
  \emph{Interactions of reggeized gluons in the Mobius representation},
  \emph{Nucl.\ Phys.\ B} {\bf 706}  391 (2005)
  [hep-ph/0404110].
\bibitem{bartels1} J.Bartels, \emph{Nucl. Phys. B} {\bf 175} 365  (1980).
\bibitem{jaro}T. Jaroszewicz, \emph{Acta Phys. Polon. B} {\bf 11}  965 (1980).
\bibitem{kwicz} J. KwieciŽnski and M. Prasza lowicz, \emph{Phys. Lett. B} {\bf 94} 413  (1980).
\bibitem{lip} L.N.Lipatov, \emph{JETP Lett.} {\bf 59} 596  (1994).
\bibitem{jw} J.Wosiek and R.A.Janik, \emph{Phys. Rev. Lett.} {\bf 79}  2935 (1997).
\bibitem{blv}
  J.~Bartels, L.~N.~Lipatov and G.~P.~Vacca,
  \emph{A New odderon solution in perturbative QCD},
  \emph{Phys.\ Lett.\ B} {\bf 477}  178 (2000).
  [hep-ph/9912423].
\bibitem{kovch1} Y. V. Kovchegov, L. Szymanowski and S. Wallon, \emph{Phys. Lett. B} {\bf 586}  267 (2004).
\bibitem{hatta}
  Y.~Hatta, E.~Iancu, K.~Itakura and L.~McLerran,
\emph{Odderon in the color glass condensate},
  \emph{Nucl.\ Phys.\ A} {\bf 760}  172 (2005)
  [hep-ph/0501171].
\bibitem{motyka} L. Motyka, \emph{Phys. Lett. B} {\bf  637}  185 (2006), [hep-ph/0509270].
\bibitem{barew} J. Bartels, C. Ewerz. \emph{JHEP} {\bf 9909}  026 (1999), [hep-ph/9908454].
\bibitem{bar2}
  J.~Bartels, M.~A.~Braun, D.~Colferai and G.~P.~Vacca,
\emph{Diffractive eta(c) photoproduction and electroproduction with the perturbative QCD odderon},
  \emph{Eur.\ Phys.\ J.\ C} {\bf 20}  323 (2001).
  [hep-ph/0102221].

\bibitem{Bartels:2012sw}
  J.~Bartels, V.~S.~Fadin, L.~N.~Lipatov and G.~P.~Vacca,
\emph{NLO Corrections to the kernel of the BKP-equations},
  \emph{Nucl.\ Phys.\ B} {\bf 867} 827  (2013)
  [arXiv:1210.0797 [hep-ph]].

\bibitem{Bartels:2013yga}
  J.~Bartels and G.~P.~Vacca,
\emph{Generalized Bootstrap Equations and possible implications for the NLO Odderon},
  \emph{Eur.\ Phys.\ J.\ C} {\bf 73} 2602  (2013)
  [arXiv:1307.3985 [hep-th]].

\bibitem{bar3}
  J.~Bartels, M.~A.~Braun and G.~P.~Vacca,
\emph{The Process $\gamma^*$ + p ---> $\eta_c$ + X: A Test for the perturbative QCD odderon},
  \emph{Eur.\ Phys.\ J.\ C} {\bf 33}  511 (2004)
  [hep-ph/0304160].
\bibitem{totem} G.Antchev {\it et al.} (TOTEM collab), [arxiv:1812.08610 [hep-ex]]
\bibitem{d0} V.M.Abazov {\it et al.} (D0 collab.), \emph{Phys. Rev. G} {\bf 86} 012009  (2012), [hep-ex/1206.0687]
\bibitem{late1} V. A. Khoze, A. D. Martin, and M. G. Ryskin, \emph{Phys. Rev. D} {\bf 97} no. 3, 034019,
[arXiv:1712.00325 [hep-ph]].
\bibitem{late2} V. A. Khoze, A. D. Martin, and M. G. Ryskin, \emph{Eur. Phys. J. C} {\bf 74} no. 2, (2014) 2756  (2018),
[arXiv:1312.3851 [hep-ph]].
\bibitem{late3}  E. Martynov and B. Nicolescu, \emph{Phys. Lett. B} {\bf 778} (2018) 414--–418, arXiv:1711.03288 [hep-ph].
\bibitem{Abramovsky:1973fm}
  V.~A.~Abramovsky, V.~N.~Gribov and O.~V.~Kancheli,
\emph{Character of Inclusive Spectra and Fluctuations Produced in Inelastic Processes by Multi - Pomeron Exchange},
  \emph{Yad.\ Fiz.}  {\bf 18}  595 (1973),
  [ \emph{Sov.\ J.\ Nucl.\ Phys.}  {\bf 18} 308 (1974) ].
\bibitem{Bartels:2005wa}
  J.~Bartels, M.~Salvadore and G.~P.~Vacca,
\emph{AGK cutting rules and multiple scattering in hadronic collisions},
  \emph{Eur.\ Phys.\ J.\ C} {\bf 42}  53 (2005).
  [hep-ph/0503049].

\bibitem{Bartels:2004ef}
  J.~Bartels, L.~N.~Lipatov and G.~P.~Vacca,
\emph{Interactions of reggeized gluons in the Mobius representation},
  \emph{Nucl.\ Phys.\ B} {\bf 706}  391 (2005).
  [hep-ph/0404110].

\bibitem{Bartels:2005ji}
  J.~Bartels, L.~N.~Lipatov, M.~Salvadore and G.~P.~Vacca,
\emph{Deformed spectral representation of the BFKL kernel and the bootstrap for gluon reggeization},
  \emph{Nucl.\ Phys.\ B} {\bf 726} 53  (2005).
  [hep-ph/0506235].

      \bibitem{Braun:1997nu}
  M.~A.~Braun and G.~P.~Vacca,
\emph{Triple pomeron vertex in the limit N(c) ---> infinity},
  \emph{Eur.\ Phys.\ J.\ C} {\bf 6} 147  (1999).
  [hep-ph/9711486].

\bibitem{Bartels:2015gou}
  J.~Bartels, C.~Contreras and G.~P.~Vacca,
\emph{Could reggeon field theory be an effective theory for QCD in the Regge limit?},
  \emph{JHEP} {\bf 1603}  201 (2016)
  [arXiv:1512.07182 [hep-th]].

\bibitem{Bartels:2016ecw}
  J.~Bartels, C.~Contreras and G.~P.~Vacca,
\emph{Pomeron - Odderon interactions in a reggeon field theory},
  \emph{Phys.\ Rev.\ D} {\bf 95}  no.1,  014013 (2017)
  [arXiv:1608.08836 [hep-th]].

  \bibitem{lip1} L.N.Lipatov, \emph{Sov. Phys. JETP} {\bf 63}  904 (1986).



\bibitem{kowalski1} H,Kowalski. L.N.Lipatov, D.S.Ross, \emph{Eur. Phys. J.} {\bf C 74}  no. 6, 2919 (2014)
\bibitem{kowalski2} H,Kowalski. L.N.Lipatov, D.S.Ross, \emph{Eur. Phys. J.} {\bf C 76} 1--23 (2016) 
\bibitem{kowalski3} H,Kowalski. L.N.Lipatov, D.S.Ross, G.Watt, \emph{Eur. Phys. J.} {\bf C 70}  983 (2010)

\bibitem{Bartels:2018pin}
  J.~Bartels, C.~Contreras and G.~P.~Vacca,
\emph{A functional RG approach for the BFKL Pomeron},
  \emph{JHEP} {\bf 1901}  004 (2019)
  [arXiv:1808.07517 [hep-ph]].

\bibitem{brarun} M.A.Braun, Phys. Lett., {\bf B 345} (1995) 155.


\bibitem{Braun:1998zj}
  M.~Braun and G.~P.~Vacca,
\emph{The 2nd order corrections to the interaction of two Reggeized gluons from the bootstrap},
  \emph{Phys.\ Lett.\ B} {\bf 454} 319  (1999)
  [hep-ph/9810454].

\bibitem{Braun:1999uz}
  M.~Braun and G.~P.~Vacca,
\emph{The Bootstrap for impact factors and the gluon wave function},
  \emph{Phys.\ Lett.\ B} {\bf 477}  156 (2000)
  [hep-ph/9910432].

\bibitem{Fadin:2002hz}
  V.~S.~Fadin and A.~Papa,
\emph{A Proof of fulfillment of the strong bootstrap condition},
  \emph{Nucl.\ Phys.\ B} {\bf 640} 309  (2002)
  [hep-ph/0206079].

\bibitem{brarun1} M.A.Braun, \emph{Eur. Phys. J. C} {\bf 53} 59  (2008).

\bibitem{Vacca:2000bk}
  G.~P.~Vacca,
\emph{Properties of a family of n reggeized gluon states in multicolor QCD},
  \emph{Phys.\ Lett.\ B} {\bf 489}  337 (2000)
  [hep-ph/0007067].

\bibitem{bvv}
  M.~Braun, G.~P.~Vacca and G.~Venturi,
\emph{Properties of the hard pomeron with a running coupling constant and the high-energy scattering},
  \emph{Phys.\ Lett.\ B} {\bf 388}  823 (1996)
  [hep-ph/9605304].

\bibitem{Bartels:2019qho}
  J.~Bartels, C.~Contreras and G.~P.~Vacca,
\emph{The Odderon in QCD with running coupling},
  arXiv:1910.04588 [hep-th].

\end{thebibliography}


\printindex

\end{document}